\documentclass[sigplan,10pt]{acmart}
\renewcommand\footnotetextcopyrightpermission[1]{}
\def\BibTeX{{\rm B\kern-.05em{\sc i\kern-.025em b}\kern-.08emT\kern-.1667em\lower.7ex\hbox{E}\kern-.125emX}}

\usepackage{formatting/shortcuts}

\begin{document}

\title[\keystone]{\keystone: An Open Framework for Architecting TEEs}

\author{Dayeol Lee}
\email{dayeol@berkeley.edu}
\affiliation{\institution{UC Berkeley}}

\author{David Kohlbrenner}
\email{dkohlbre@berkeley.edu}
\affiliation{\institution{UC Berkeley}}

\author{Shweta Shinde}
\email{shwetas@berkeley.edu}
\affiliation{\institution{UC Berkeley}}

\author{Krste Asanovi\'{c}}
\email{krste@berkeley.edu}
\affiliation{\institution{UC Berkeley}}

\author{Dawn Song}
\email{dawnsong@berkeley.edu}
\affiliation{\institution{UC Berkeley}}

\begin{abstract}
Trusted execution environments (TEEs) are being used in all the
devices from embedded sensors to cloud servers and encompass a range
of cost, power constraints, and security threat model choices. On the
other hand, each of the current vendor-specific TEEs makes a fixed set
of trade-offs with little room for customization. We present
\codename---the first open-source framework for building customized
TEEs. \codename uses simple abstractions provided by the hardware such
as memory isolation and a programmable layer underneath untrusted
components (e.g., OS). We build reusable TEE core primitives from
these abstractions while allowing platform-specific modifications and
application features. We showcase how \codename-based TEEs run on
unmodified \riscv hardware and demonstrate the strengths of our design
in terms of security, TCB size, execution of a range of benchmarks,
applications, kernels, and deployment models.
\end{abstract}
\maketitle
\thispagestyle{empty}
\section{Introduction}
\label{sec:introduction}

The last decade has seen the proliferation of trusted execution
environments (TEEs) to protect sensitive code and data. All major CPU
vendors have rolled out their TEEs (e.g., ARM TrustZone, Intel
SGX, and AMD SEV) to create a secure execution environment, commonly
referred to as an {\em enclave} ~\cite{sgx, trustzone, amd-sev}. On 
the consumer end, TEEs are now being used for secure cloud
services~\cite{haven, scone}, databases~\cite{enclavedb}, big data
computations~\cite{vc3, m2r, securekeeper}, secure
banking~\cite{droidvault}, blockchain consensus
protocols~\cite{teechain, proof-of-luck, poet}, smart
contracts~\cite{towncrier, ekiden, tesseract}, machine
learning~\cite{sgx-ml, privado}, network middleboxes~\cite{endbox,
vif}, and so on. These use-cases have diverse deployment environments
ranging from cloud servers, client devices, mobile phones, ISPs, IoT
devices, sensors, and hardware tokens.
{\let\thefootnote\relax\footnote{{\codename is available at
\keystoneurl }}}

Unfortunately, each vendor TEE enables only a small portion of the
possible design space across threat models, hardware requirements,
resource management, porting effort, and feature compatibility. When a
cloud provider or software developer chooses a target hardware
platform they are locked into the respective TEE design limitations
regardless of their actual application needs. Constraints breed
creativity, giving rise to significant research effort in working
around these limits. For example, Intel SGXv1~\cite{sgx} requires 
statically sized enclaves, lacks secure I/O and syscall support, and
is vulnerable to significant side-channels~\cite{sgx-explained}. Thus,
to execute arbitrary applications, the systems built on SGXv1 have
inflated the Trusted Computing Base (TCB) and are forced to implement
complex workarounds~\cite{haven, enclavedb, scone, graphene-sgx,
panoply}. As only Intel can make changes to the inherent design
trade-offs in SGX, users have to wait for changes like dynamic
resizing of enclave virtual memory in the proposed
SGXv2~\cite{sgx-dyn-mem}. Unsurprisingly, these and other similar 
restriction have led to a proliferation of TEE re-implementations on
other ISAs (e.g., OpenSPARC~\cite{sp, bastion}, RISC-V~\cite{sanctum,
timberv}). However, each such redesign requires considerable effort
and only serves to create another fixed design point.

In this paper, we advocate that the hardware must provide security
{\em primitives} and not point-wise solutions. We can draw an analogy
with the move from traditional networking solutions to Software
Defined Networking (SDN), where exposing the packet forwarding
primitives to the software has led to far more use-cases and research
innovation. We believe a similar paradigm shift in TEEs will pave the
way for effortless customization. It will allow features and the
security model to be tuned for each hardware platform and use-case
from a common base. This motivates the need for {\em \programmable
TEEs} to provide a better interface between entities that create the
hardware, operate it, and develop applications. \Programmable TEEs
promise quick prototyping, a shorter turn-around time for fixes,
adaptation to threat models, and use-case specific deployment.

The first challenge in realizing this vision is the lack of a
programmable trusted layer below the {\em untrusted} OS. Hypervisor
solutions result in a trusted layer with a mix of security and
virtualization responsibilities, unnecessarily complicating the most
critical component. Firmware and micro-code are not programmable to a
degree that satisfies this requirement. Second, previous attempts at
re-using hardware isolation to provide TEE guarantees do not capture 
the right notion of customization. Specifically, these solutions (e.g.
Komodo~\cite{komodo}, \sel~\cite{sel4}) rely on the underlying
hardware not only for a separation mechanism but also for pre-made
decision about the boundary between what is trusted and untrusted.
They use a verified trusted software layer for customization on top of
this hardware~\cite{komodo}. To overcome these challenges, our insight
is to identify the appropriate hardware memory isolation primitives
and then build a modular software-programmable layer with the
appropriate privileges 

To this end, we propose \codename---the first open-source framework
for building \programmable TEEs. We build \codename on unmodified
\riscv using its standard specifications~\cite{riscv-priv-110} for
physical memory protection (PMP)---a primitive which allows the
programmable machine mode underneath the OS in \riscv to specify
arbitrary protections on physical memory regions. We use this machine
mode to execute a trusted {\em security monitor (SM)} to provide
security boundaries without needing to perform any resource
management. More importantly, enclave operates in its own
isolated physical memory region and has its own {\em
runtime} (\rt) component which executes in supervisor mode and manages
the virtual memory of the enclave. With this novel design, any
enclave-specific functionality can be implemented cleanly by its \rt
while the \sm manages hardware-enforced guarantees. This allows the
enclaves to specify and include only the necessary components. The \rt
implements the functionality, communicates with the \sm, mediates
communication with the host via shared memory, and services {\em
enclave user-code application (\eapp)}.

Our choice of \riscv and the logical separation between \sm and \rt
allows hardware manufacturers, cloud providers, and application
developers to configure various design choices such as TCB, threat
models, workloads, and TEE functionality via compile-time plugins.
Specifically, \codename's \sm uses hardware primitives to provide
in-built support for TEE guarantees such as secure boot, memory
isolation, and attestation. The \rt then provides functionality
plugins for system call interfaces, standard $\tt{libc}$ support,
in-enclave virtual memory management, self-paging, and more inside the
enclave. For strengthening the security, our \sm leverages a highly
configurable cache controller to transparently enable defenses against
physical adversaries and cache side-channels.

We build \codename, the \sm, and an \rt (called \eyrie) which together
allow enclave-bound user applications to selectively use all the above
features. We demonstrate an off-the-shelf microkernel
(\sel~\cite{sel4}) as an alternative \rt. We extensively benchmark
\codename on $4$ suites with varying workloads: RV8, IOZone, CoreMark,
and Beebs. We showcase use-case studies where \codename can be used
for secure machine learning (Torch and FANN frameworks) and
cryptographic tasks (sodium library) on embedded devices and cloud
servers. Lastly, we test \codename on different \riscv systems: the
\hifivefull, $3$ in-order cores and $1$ out-of-order core via FPGA,
and a QEMU emulation---all without modification. \codename is
open-source and available online~\cite{anon-code-release}.

\paragraph{Contributions.} We make the following contributions:
\begin{itemize}

\item{\em Need for \Programmable TEEs.}
We define a new paradigm wherein the hardware
manufacturer, hardware operator, and the enclave programmer can 
tailor the TEE design.

\item{\em \codename Framework.}
We present the first framework to build and instantiate \programmable
TEEs. Our principled way of ensuring modularity in \codename allows us
to customize the design dimensions of TEE instances as per the
requirements.

\item {\em Open-source Implementation.}
We demonstrate the advantages of \codename TEEs which can minimize the
TCB, adapt to threat models, use hardware features, handle workloads,
and provide rich functionality without any micro-architectural changes.
Total TCB of a \codename enclaved application is between $12$-$15$~K
lines of code (\loc), of which the \sm consists of only $1.6$~K\loc
added by \codename.

\item {\em Benchmarking \& Real-world Applications.}
We evaluate \codename on $4$ benchmarks: CoreMark, Beebs, and RV8
($<1\%$ overhead), and IOZone ($40\%$). We demonstrate real-world
machine learning workloads with Torch in \eyrie ($7.35\%$), FANN
($0.36\%$) with \sel, and a \codename-native secure remote computation
application. Finally, we demonstrate enclave defenses for adversaries
with physical access via memory encryption and a cache side-channel
defense.

\end{itemize}
\section{\Programmable TEEs}
\label{sec:problem}
\begin{figure}
\includegraphics[width=0.99\columnwidth]{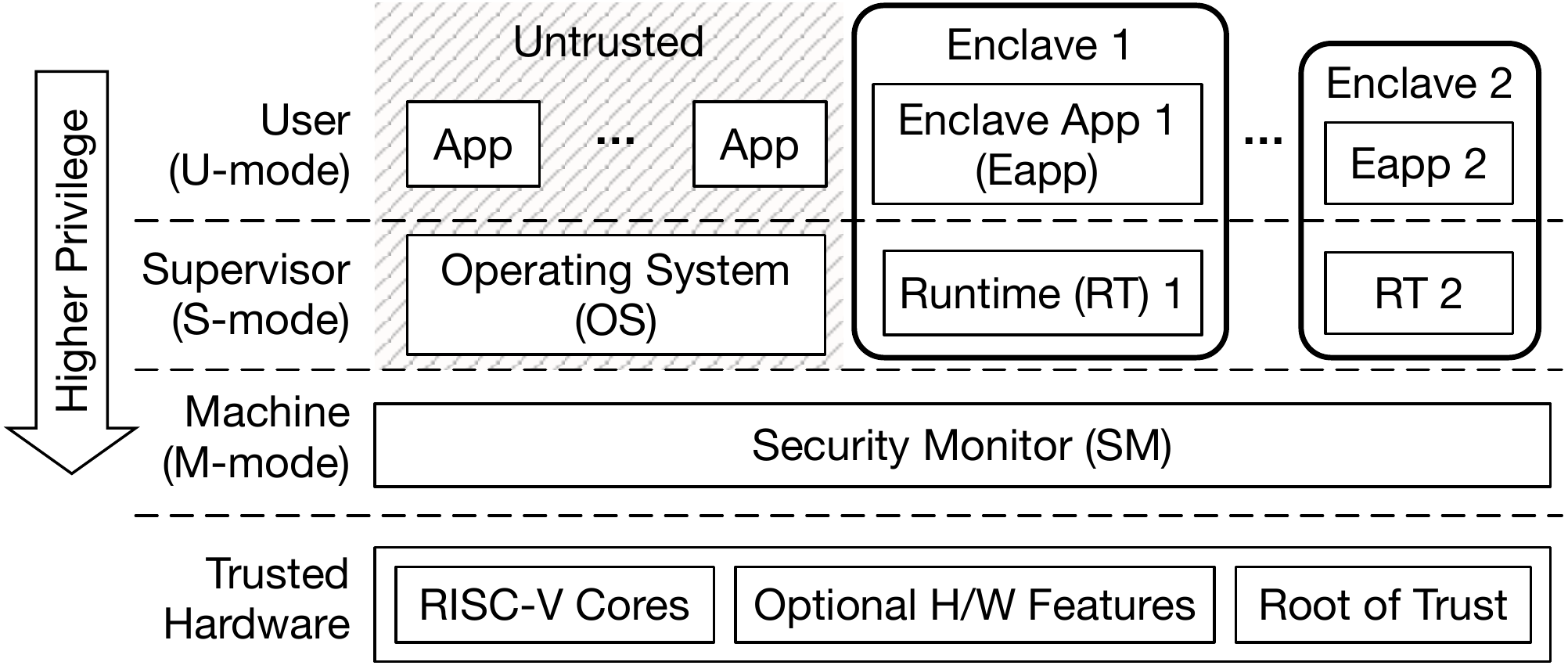}
\caption{
\codename overview of the system setup and privilege levels. Includes
components such as untrusted host processes, the untrusted OS, the
security monitor, and multiple enclaves (each with a runtime and an
\eapp).}
\label{fig:keystone_overview}
\end{figure}

We motivate the need for \programmable TEEs which can adapt to 
various real deployment scenarios, introduce our key actors, and
define our goals.

\subsection{Need for a New Paradigm.}
Current widely-used TEE systems cater to specific and valuable 
use-cases but occupy only a small part of the wide design space (see
Table~\ref{tab:comparison}). Consider the case of heavy server
workload (databases, ML inference, etc.) running in an untrusted cloud
environment. One option is an SGX-based solution which has a large
software stack~\cite{haven, graphene-sgx, scone} to extend the
supported features. On the other hand, a SEV-based solution requires a
complete virtualization stack. If one wants additional defenses
against side-channels, it adds further user-space software mechanisms
for both cases. If we consider edge-sensors or IoT applications, the
available solutions are TrustZone based. While more flexible than SGX
or SEV, TrustZone supports only a single hardware-enforced isolated
domain called the Secure World. Any further isolation needs
multiplexing between applications within one domain with
software-based Secure World OS solutions~\cite{optee}. Thus,
irrespective of the TEE, developers often compromise their design
requirements (e.g., resort to a large TCB solution, limited privilege
domains) or take on the onus of building their custom design.

\paragraph{Hardware-software co-design.}
\codename builds on simple yet elegant primitives. We leverage
multiple well-defined and existing hardware features while allowing
the software layers to manage the complex responsibilities. Previous
works explore the spectrum of tasks the hardware and software should
do. For instance, \sel~\cite{sel4} uses a single isolation domain
exposed by the hardware and performs almost all resource management,
isolation, and security enforcement in a verified software layer.
Komodo uses the two isolation domains provided by ARM TrustZone to
execute multiple isolated domains enforced by formally verified
software. In \codename, we expect arbitrary hardware-isolated domains
whose boundaries can be dynamically configured in the software. With
this novel choice, \codename provides guarantees with minimum trusted
software while allowing the isolated regions to manage themselves.

\begin{table*}[]
\centering
\resizebox{0.99\textwidth}{!}{%
\begin{tabular}{clcccccccccc}
\hline
\textbf{} & \multicolumn{1}{c}{\textbf{System}} & 
\begin{tabular}{@{}c@{}}\textbf{Software}\\ \textbf{Adversary} \\ \textbf{Protection}\end{tabular} & 
\begin{tabular}{@{}c@{}}\textbf{Physical} \\ \textbf{Adversary} \\ \textbf{Protection}\end{tabular} & 
\begin{tabular}{@{}c@{}}\textbf{Side Ch}\\ \textbf{Adversary} \\ \textbf{Protection}\end{tabular} & 
\begin{tabular}{@{}c@{}}\textbf{Control Ch} \\ \textbf{Adversary} \\ \textbf{Protection}\end{tabular} & 
\begin{tabular}{@{}c@{}}\textbf{Low} \\ \textbf{SW} \\ \textbf{TCB}\end{tabular} & 
\begin{tabular}{@{}c@{}}\textbf{No} \\ \textbf{HW} \\ \textbf{Mods}\end{tabular} & 
\begin{tabular}{@{}c@{}}\textbf{Flexible} \\ \textbf{Resource} \\ \textbf{Mgmt}\end{tabular} & 
\begin{tabular}{@{}c@{}}\textbf{Range} \\ \textbf{of Apps} \\ \textbf{Supported}\end{tabular} & 
\begin{tabular}{@{}c@{}}\textbf{Supports} \\ \textbf{High} \\ \textbf{Expr}\end{tabular} & 
\begin{tabular}{@{}c@{}}\textbf{Low} \\ \textbf{Porting} \\ \textbf{Effort}\end{tabular} \\ \hline \addlinespace[0.2cm]
 & SGX~\cite{sgx} & \compfull & \compfull & \compnone & \compnone & \compfull & \compnone & \compnone & \compnone & \compnone & \compnone \\
 & Haven~\cite{haven} & \compfull & \compfull & \compnone & \compnone & \compnone & \compfull & \comppart & \comppart & \compfull & \compfull \\
 & Graphene-SGX~\cite{graphene-sgx} & \compfull & \compfull & \compnone & \compnone & \compnone & \compfull & \comppart & \comppart & \compfull & \compfull \\
 & Scone~\cite{scone} & \compfull & \compfull & \compnone & \compnone & \comppart & \compfull & \compnone & \comppart & \compfull & \comppart \\
\multirow{-5}{*}{\ver{Intel}} & Varys~\cite{varys} & \compfull & \compfull & \compfull & \comppart & \comppart & \compnone & \compnone & \comppart & \compfull & \comppart \\ \hline \addlinespace[0.15cm]
 & TrustZone~\cite{trustzone} & \compfull & \compnone & \compnone & \compnone & \compnone & \compnone & \compfull & \compfull & \compfull & \compnone \\
 & Komodo~\cite{komodo} & \compfull & \compfull & \compnone & \compfull & \comppart & \compfull & \comppart & \compfull & \comppart & \compfull \\
 & OP-TEE~\cite{optee} & \compfull & \compfull & \compnone & \compnone & \compnone & \compfull & \compfull & \compfull & \compfull & \compfull \\
\multirow{-4}{*}{\ver{ARM}} & Sanctuary~\cite{sanctuary} & \compfull & \compnone & \comppart & \compnone & \compnone & \compfull & \comppart & \compfull & \compfull & \compfull \\ \hline \addlinespace[0.15cm]
 & SEV~\cite{amd-sev} & \comppart & \comppart & \compnone & \compnone & \compfull & \compnone & \compfull & \compfull & \compfull & \compfull \\
\multirow{-2}{*}{\ver{AMD}} & SEV-ES~\cite{amd-sev-es} & \compfull & \compfull & \compnone & \compnone & \compfull & \compnone & \compfull & \compfull & \compfull & \compfull \\ \hline \addlinespace[0.15cm]
 & Sanctum~\cite{sanctum} & \compfull & \compnone & \compfull & \compfull & \comppart & \compnone & \comppart & \comppart & \comppart & \compfull \\
 & TIMBER-V~\cite{timberv} & \compfull & \compfull & \compnone & \compnone & \compfull & \compnone & \compfull & \comppart & \compfull & \comppart \\
 & MultiZone~\cite{multizone} & \compfull & \compfull & \compfull & \compfull & \comppart & \compfull & \compnone & \compnone & \compnone & \comppart \\
\multirow{-4}{*}{\ver{RISC-V}} & \cellcolor[HTML]{EFEFEF}\textbf{\codename (This paper)} & \cellcolor[HTML]{EFEFEF}\compfull & \cellcolor[HTML]{EFEFEF}\compfull & \cellcolor[HTML]{EFEFEF}\compfull & \cellcolor[HTML]{EFEFEF}\compfull & \cellcolor[HTML]{EFEFEF}\comppart & \cellcolor[HTML]{EFEFEF}\compfull & \cellcolor[HTML]{EFEFEF}\compfull & \cellcolor[HTML]{EFEFEF}\compfull & \cellcolor[HTML]{EFEFEF}\comppart & \cellcolor[HTML]{EFEFEF}\compfull \\ \hline
\end{tabular}%
}
\vspace{10pt}
\caption{Trade-offs in existing TEEs and extensions.
\textbf{C1-2}: Category and specific systems.
\compfull, \comppart, \compnone ~represent best to least adherence to
the property in the column respectively.
\textbf{C3-6}: resilience against software adversary, hardware
adversary, side-channel adversary, control-channel adversary
respectively; the symbols indicates complete protection,
only confidentiality protection, and no protection respectively. 
\textbf{C7}: zero, thousands of \loc, millions of \loc of TCB.
\textbf{C8}: zero or non-zero hardware / micro-architectural modifications.
\textbf{C9}: 
allows the enclave to do self resource management with complete, partial, or no flexibility.
\textbf{C10}: the range of applications supported are maximum;
specific class; only written from scratch. 
\textbf{C11}: expressiveness includes forking, multi-threading,
syscalls, shared memory; partial; none of these.
\textbf{C12}: developer effort for porting is zero (unmodified
binaries); partial (compiling and / or configuration files);
substantial (significant re-writing).} 
\label{tab:comparison}
\end{table*}

\paragraph{\Programmable TEEs.} 
We define a new paradigm---{\em \programmable TEEs}---to allow
multiple stakeholders to customize a TEE. The hardware manufacturer
only provides basic primitives. Realizing a specific TEE instance
involves the platform provider's choice of the hardware interface, the
trust model, and the enclave programmer's feature requirements. The
entities offload their choices to a framework which plugs in required
components and composes them to instantiate the required TEE. With
this, we bridge the existing gaps in the TEE design space, allow
researchers to independently explore design trade-offs without
significant development effort, and encourage rapid response to
vulnerabilities and new feature requirements.

\paragraph{Trusted Hardware Requirements.}
\codename requires no changes to CPU cores, memory controllers, etc. A
secure hardware platform supporting \codename has several feature
requirements: a device-specific secret key visible only to the trusted
boot process, a hardware source of randomness, and a trusted boot
process. Key provisioning~\cite{sgx-seal} is an orthogonal problem.
For this paper, we assume a simple manufacturer provisioned key.

\subsection{Entities in TEE Lifecycle}
\label{sec:actors}

We define six logical entities in \programmable TEEs:
\\
\emph{\bf \designer} designs the hardware; designs or modifies
existing SoCs, IP blocks and interactions.
\\
\emph{\bf \manufacturer} fabricates the hardware; assembles them as
per the pre-defined design.
\\
\emph{\bf \provider} purchases manufactured hardware; operates
the hardware; makes it available for use to its customers; configures
the \sm.
\\
\emph{\bf \codenamedev} develops \codename software components
including \sm, \rt, and \eapps; we refer to the programmers who 
develop these specific components as \smdev, \rtdev, \eappdev
respectively. 
\\
\emph{\bf \codenameuser} chooses a \codename configuration of \rt and
an \eapp. They instantiate a TEE which can execute on their choice of
hardware provisioned by the \provider.
\\
\emph{\bf \eappuser} interacts with the \eapp executing in an enclave
on the TEE instantiated using \codename. 
\\
We define these fine-granularity roles for customization options in
\codename. In real-world deployments, a single entity can perform
multiple roles. For example, consider the scenario where Acme Corp.
hosts their website on an Apache webserver executing on Bar Corp.
manufactured hardware in a \codename-based enclave hosted on Cloud
Corp. cloud service. In this scenario, Bar will be the \designer and
\manufacturer; Cloud will be a \provider and can be an \rtdev and
\smdev; Apache developers will be \eappdev; Acme Corp. will be
\codename user, and; the person who uses the website will be the
\eappuser.
\section{\codename Overview}
\label{sec:approach}

We use \riscv to design and build \codename. \riscv is an open ISA
with multiple open-source core implementations~\cite{rocket, boom}. It
currently supports up to three privilege modes: U-mode (user) for
normal user-space processes, S-mode (supervisor) for the kernel, and
M-mode (machine) which can directly access physical resources (e.g.,
interrupts, memory, devices).

\subsection{Design Principles}

We design \programmable TEEs with maximum degrees of freedom and
minimum effort using the following principles.

\paragraph{Leverage programmable layer and isolation primitives below
the untrusted code.}
We design a \securitymonitor (\sm) to enforce TEE guarantees on the
platform using three properties of M-mode: (a) it is programmable by
platform providers and meets our needs for a minimal highest privilege
mode. (b) It controls hardware delegation of the interrupts and
exceptions in the system. All the lower privilege modes can receive
exceptions or use CPU cycles only when the M-mode allows it. (c)
Physical memory protection (PMP), a feature of the \riscv Privileged
ISA~\cite{riscv-priv-110}, enables the M-mode to enforce simple access
policies to physical memory, allowing isolation of memory regions at
runtime or disabling access to memory-mapped control features.

\paragraph{Decouple the resource management and security checks.} The
\sm enforces security policies with minimal code at the highest
privilege. It has few non-security responsibilities. This lowers its
TCB and allows it to present clean abstractions. Our S-mode \runtime
(\rt) and U-mode enclave application (\eapp) are mutually trusting,
reside in enclave address space, and are isolated from the untrusted
OS or other user applications. The \rt manages the lifecycle of the
user code executing in the enclave, manages memory, services syscalls,
etc. For communication with the \sm, the \rt uses a limited set of API
functions via the \riscv supervisor binary interface (SBI) to exit or
pause the enclave (Table~\ref{table:sm-api}) as well as request \sm
operations on behalf of the \eapp (e.g., attestation). Each enclave
instance may choose its own \rt which is never shared with any other
enclave.

\paragraph{Design modular layers.}
\codename uses modularity (\sm, \rt, \eapp) to support a variety of
workloads. It frees \provider{}s and \codenamedev{}s from retrofitting
their requirements and legacy applications into an existing TEE
design. Each layer is independent, provides a security-aware
abstraction to the layers above it, enforces guarantees which can be
easily checked by the lower layers, and is compatible with existing
notions of privilege.

\paragraph{Allow to selectively include TCB only when required.}
\codename can instantiate TEEs with minimal TCB for each of the
specific expected use-cases. The enclave programmer can further
optimize the TCB via \rt choice and \eapp libraries using existing
user/kernel privilege separation. For example, if the \eapp does not
need libc support or dynamic memory management, \codename will not
include them in the enclave. 

\subsection{Threat Model}
\label{sec:threat_model}

The \codename framework trusts the PMP specification as well as the
PMP and \riscv hardware implementation to be bug-free. The
\codenameuser trusts the \sm only after verifying if the \sm
measurement is correct, signed by trusted hardware, and has the
expected version. The \sm only trusts the hardware, the \rt trusts the
\sm, the \eapp trusts the \sm and the \rt.

\paragraph{Adaptive Threat Model.} 
\codename can operate under diverse threat models, each requiring
different defense mechanisms. For this reason, we outline all the
relevant attackers for \codename. We allow the selection of a sub-set
of these attackers based on the scenario. For example, if the user is
deploying TEEs in their own private data centers or home appliances, a
physical attacker may not be a realistic threat and \codename can be
configured to operate without physical adversary protections.

\paragraph{Attacker Models.}
\codename protects the confidentiality and integrity of all the
enclave code and data at all points during execution. We define four
classes of attackers who aim to compromise our security guarantees:
\\
A {\bf \em physical attacker} \attackerphy can intercept, modify, or
replay signals that leave the chip package. We assume that the
physical attacker does not affect the components inside the chip
package. \attackerphc is for confidentiality, \attackerphi is for
integrity.
\\
A {\bf \em software attacker} \attackersw can control host applications,
the untrusted OS, network communications, launch adversarial enclaves,
arbitrarily modify any memory not protected by the TEE, and
add/drop/replay enclave messages. 
\\
A {\bf \em side-channel attacker} \attackersc can glean information by
passively observing interactions between the trusted and the
untrusted components via the cache side-channel (\attackerscc), the
timing side-channel (\attackersct) or the control channel
(\attackerccs).
\\
A {\bf \em denial-of-service attacker} \attackerdos can take down the
enclave or the host OS. \codename allows these attacks against
enclaves as the OS can refuse services to user applications at any
time.
 
\paragraph{Scope.}
\codename currently has no meaningful mechanisms to protect against
speculative execution attacks~\cite{spectre, foreshadow, zombieload,
ridl, fallout, boom-speculative}. We recommend \codename users to
avoid deployments on out-of-order cores. Existing and future defenses
against this class of attacks can be retrofitted into
\codename~\cite{mi6, invisispec}. \codename does not natively protect
the enclave or the \sm against timing side-channel attacks. \sm, \rt,
and \eapp programmers should use existing software solutions to mask
timing channels~\cite{time-sc-survey} and \manufacturer{}s can supply
timing side-channel resistant hardware~\cite{dawg}. The \sm exposes a
limited API to the host OS and the enclave. We do not provide
non-interference guarantees for this API~\cite{komodo, nickel}.
Similarly, the \rt can optionally perform untrusted system calls into
the host OS. We assume that the \rt and the \eapp have sufficient
checks in place to detect Iago attacks via this untrusted
interface~\cite{besfs, cosmix, graphene-sgx}, and the \sm, \rt, and
\eapp are bug-free.
\section{\codename Internals}
\label{sec:internals}

We discuss the \codename design details and the enclave lifecycle. Our
enclave consists of a kernel-like component in the S-mode (the
runtime---\rt), and an application in the U-mode (the enclave
application---\eapp). The \rt and the host OS share a dedicated buffer
that is accessible only to specific \rt functions and the host.

\begin{table}[]
\centering
\resizebox{0.45\textwidth}{!}{%
\begin{tabular}{@{}lll@{}}
\toprule
\textbf{Caller}     & \textbf{SM API} & \textbf{Description}              \\ \midrule
\multirow{5}{*}{OS} & create          & Validate, and measure the enclave \\
                    & run             & Start enclave and boot RT         \\
                    & resume          & Resume enclave execution          \\
                    & destroy         & Release enclave memory            \\
                    \midrule
\multirow{4}{*}{RT} & stop            & Pause enclave execution           \\
                    & exit            & Terminate the enclave             \\
                    & attest          & Get a signed attestation report   \\
                    & random          & Get secure random values          \\ \midrule 
OS \& RT 			& plugin*         & Call SM plugins (e.g., dynamic resizing) \\
                    \bottomrule
\end{tabular}%
}
\vspace{10pt}
\caption{The SBI API functions (SBI) provided by the \codename SM. 
* indicates that the SBI is provided by an optional SM plugin.}
\vspace{-10pt}
\label{table:sm-api}
\end{table}

\subsection{\codename Memory Isolation Primitives}
\label{sec:tee_primitives}

\codename has significantly better modularity and customizability 
because of our conscious design choices. First, we expect the hardware
to provide only simple security primitives and interfaces. Second, we
assign minimum responsibilities of the trusted software components
executing at the highest privilege (e.g., bootloader, \sm). Finally,
we push bulk of the resource management logic either to the untrusted
software or the enclave.

\begin{figure}
\includegraphics[width=0.99\columnwidth]{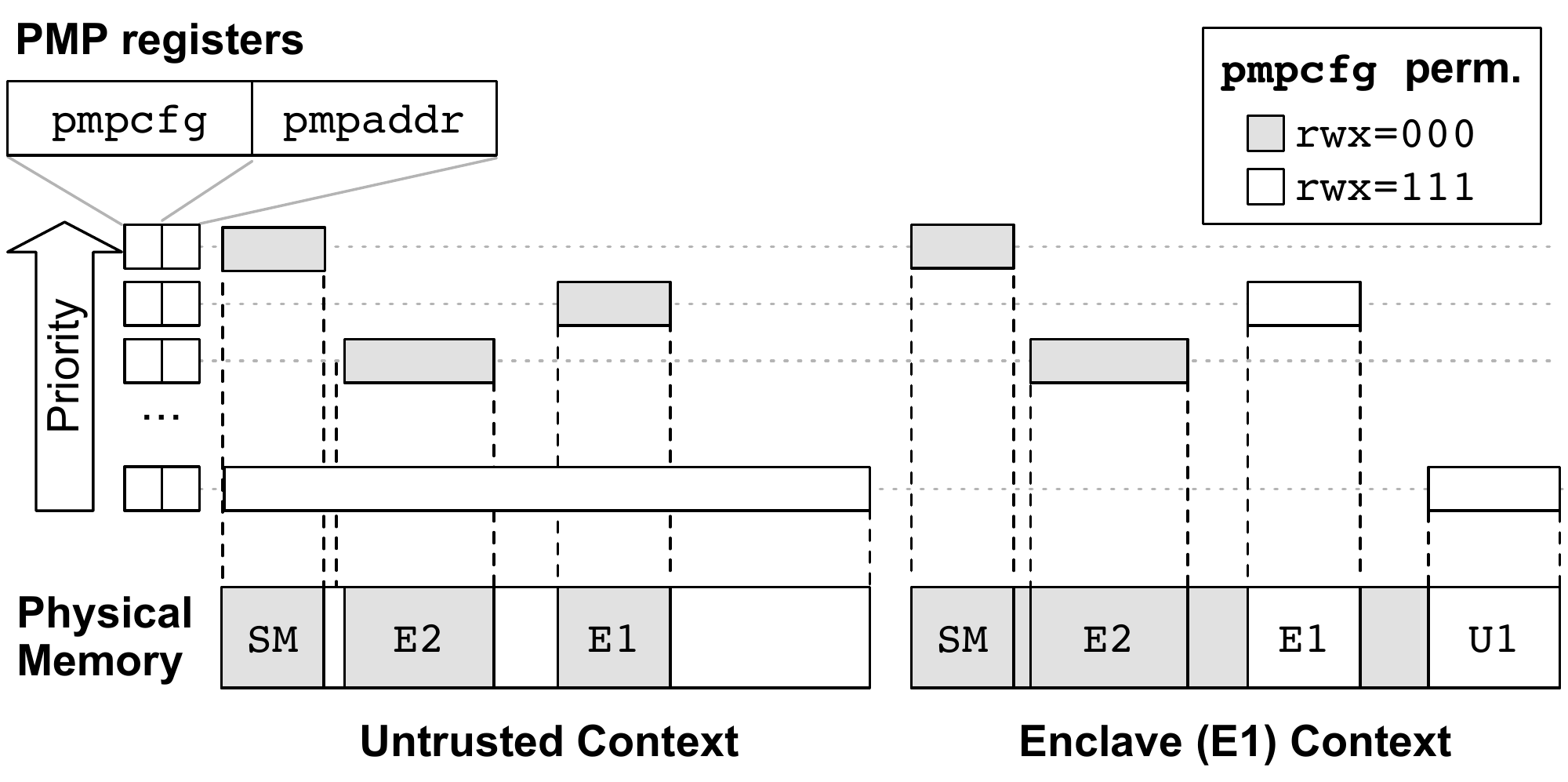}
\caption{
\keystone's use of \riscv PMP for the flexible, dynamic memory
isolation. The SM uses a few PMP entries to guard its own memory
(\texttt{SM}), enclave memory regions ($\tt{E1}$, $\tt{E2}$),  and the
untrusted buffer ($\tt{U1}$). On enclave entry, the SM will
reconfigure the PMP such that the enclave can only access its own
memory ($\tt{E1}$) and the enclave's untrusted buffer ($\tt{U1}$).
}
\vspace{-10pt}
\label{fig:pmp_full}
\end{figure}

\paragraph{Background: \riscv Physical Memory Protection.}
\codename uses physical memory protection (PMP) feature provided by
\riscv. PMP restricts the physical memory access of S-mode and U-mode
to certain regions defined via {\em PMP entries} (See
Figure~\ref{fig:pmp_full}).\footnote{\texttt{pmpaddr} and
\texttt{pmpcfg} CSRs are used to specify PMP entries} Each PMP entry
controls the U-mode and S-mode permissions to a customizable region of
physical memory.\footnote{Currently processors have up to $16$ M-mode
configurable PMP entries.} The PMP address registers encode the
address of a contiguous physical region, configuration bits specify
the r-w-x permissions for U/S-mode, and two addressing mode bits. PMP
has three addressing modes to support various sizes of regions
(arbitrary regions and power-of-two aligned regions). PMP entries are
statically prioritized with the lower-numbered PMP entries taking
priority over the higher-numbered entries. If any mode attempts to
access a physical address and it does not match any PMP address range,
the hardware does not grant any access permissions.

\paragraph{Enforcing Memory Isolation via \sm.}
PMP makes \codename memory isolation enforcement flexible for three
reasons: (a) multiple discontiguous enclave memory regions can coexist
instead of reserving one large memory region shared by all enclaves;
(b) PMP entries can cover a regions between sizes $4$ bytes or maximum
DRAM size so \codename enclaves utilize page-aligned memory with an
arbitrary size; (c) PMP entries can be dynamically reconfigured during
execution such that \codename can dynamically create a new region or
release a region to the OS.

During the SM boot, \codename configures the first PMP entry (highest
priority) to cover its own memory region (code, stack, data such as
enclave metadata and keys). The \sm disallows all access to its
memory from U-mode and S-mode using this first entry and configures
the last PMP entry (lowest priority) to cover all memory and with all
permissions enabled. The OS thus has default full permissions to all
memory regions not otherwise covered by a PMP entry.

When a host application requests the OS to create an enclave, the OS
finds an appropriate contiguous physical region. On receiving a valid
enclave creation request, the \sm protects the enclave memory by
adding a PMP entry with all permissions disabled. Since the enclave's
PMP entry has a higher priority than the OS PMP entry (the last in
Figure~\ref{fig:pmp_full}), the OS and other user processes cannot
access the enclave region. A valid request requires that enclave
regions not overlap with each other or with the \sm region.

At a CPU control-transfer to an enclave, the \sm (for the current core
only): (a) enables PMP permission bits of the relevant enclave memory
region; and (b) removes all OS PMP entry permissions to protect all
other memory from the enclave. This allows the enclave to access its
own memory and no other regions. At a CPU context-switch to
non-enclave, the \sm disables all permissions for the enclave region
and re-enables the OS PMP entry to allow default access from the OS.
Enclave PMP entries are freed on enclave destruction.

\paragraph{PMP Enforcement Across Cores.}
Each core has its own complete set of PMP entries. During enclave
creation, \codename adds a PMP entry to disallow everyone from
accessing the enclave. These changes during the creation must be
propagated to all the cores via inter-processor interrupts (IPIs). 
The \sm executing on each of the cores handles these IPIs by removing 
the access of other cores to the enclave. During the enclave
execution, changes to the PMP entries (e.g., context switches between
the enclave and the host) are local to the core executing it and need
not be propagated to the other cores. When \codename destroys or
creates an enclave, all the other cores are notified to update their
PMP entries. There are no other times when the PMPs need to be
synchronized via IPIs.

\subsection{Enclave Lifecycle}
\label{sec:lifecycle}

\begin{figure}
\includegraphics[width=\columnwidth]{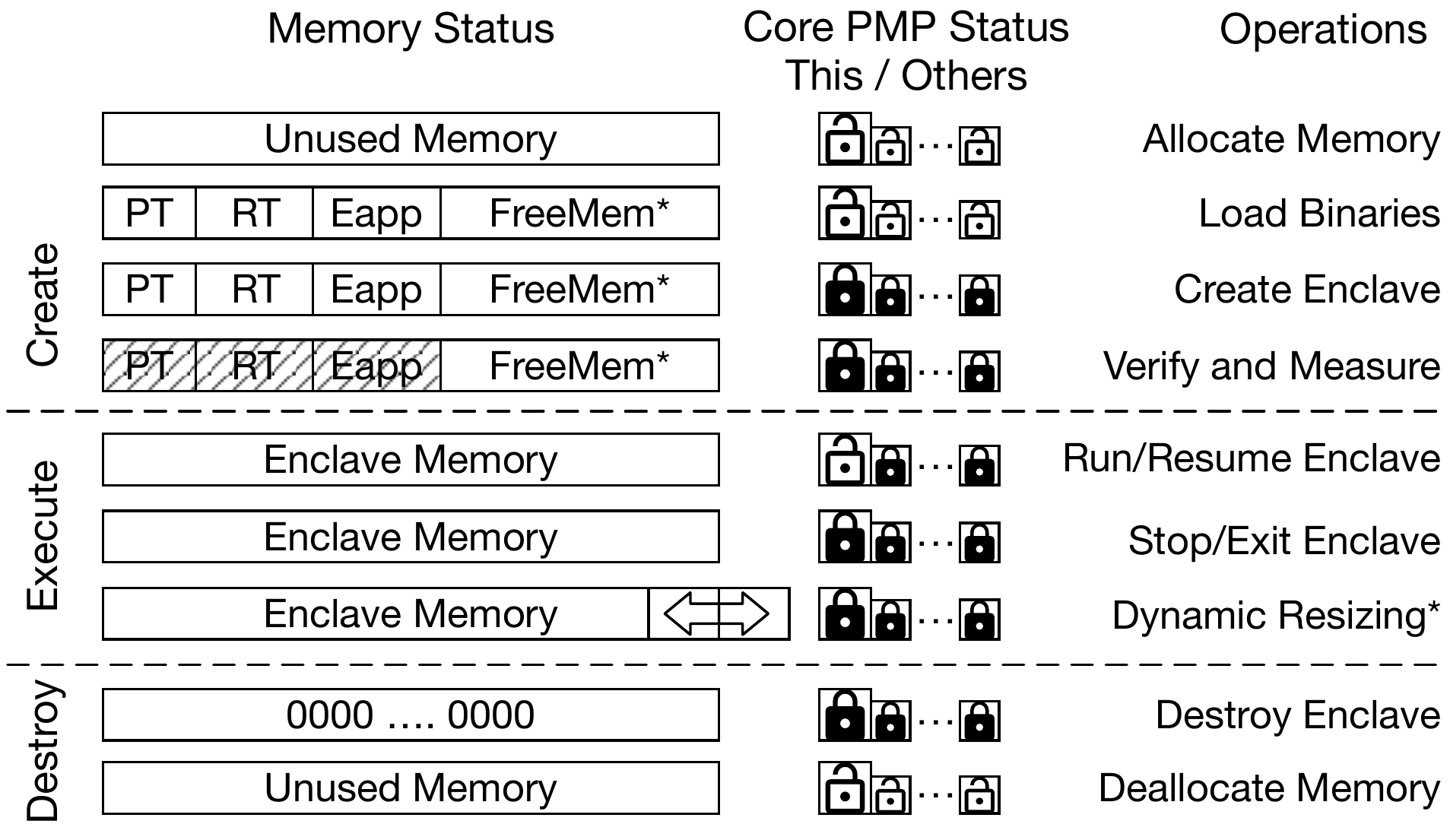}
\caption{
Lifecycle of an enclave. The enclave memory and the corresponding PMP
entry status (accessible or not) are shown per each operation. For
PMP status, \textit{This} means the PMP status of the core performing
the operation and \textit{Others} is PMP of other cores.
}
\label{fig:enclave_lifecycle}
\end{figure}

We summarize the end-to-end life cycle of a \codename enclave. 
Figure~\ref{fig:enclave_lifecycle} shows the key steps in the
enclave lifetime and the corresponding PMP changes.
\\
{\bf \em Enclave Creation.}
At creation, \codename measures the enclave memory to ensure that the
OS has loaded the enclave binaries correctly to the physical memory.
\codename uses the initial virtual memory layout for the measurement 
because the physical layout can legitimately vary (within limits)
across different executions. For this, the \sm expects the OS to
initialize the enclave page tables and allocate physical memory for
the enclave. The \sm walks the OS-provided page table and checks if
there are invalid mappings and ensures a unique virtual-to-physical
mapping. Then the \sm hashes the page contents along with the virtual
addresses and the configuration data.
\\
{\bf \em Enclave Execution.} 
The \sm sets the PMP entries and transfers the control to the enclave
entry point.
\\
{\bf \em Enclave Destruction.}
On an OS initiated tear-down, the \sm clears the enclave memory region
before returning the memory to the OS. \sm cleans and frees all the
enclave resources, PMP entries, and enclave metadata.

\begin{figure}
\center
\includegraphics[width=0.97\columnwidth]{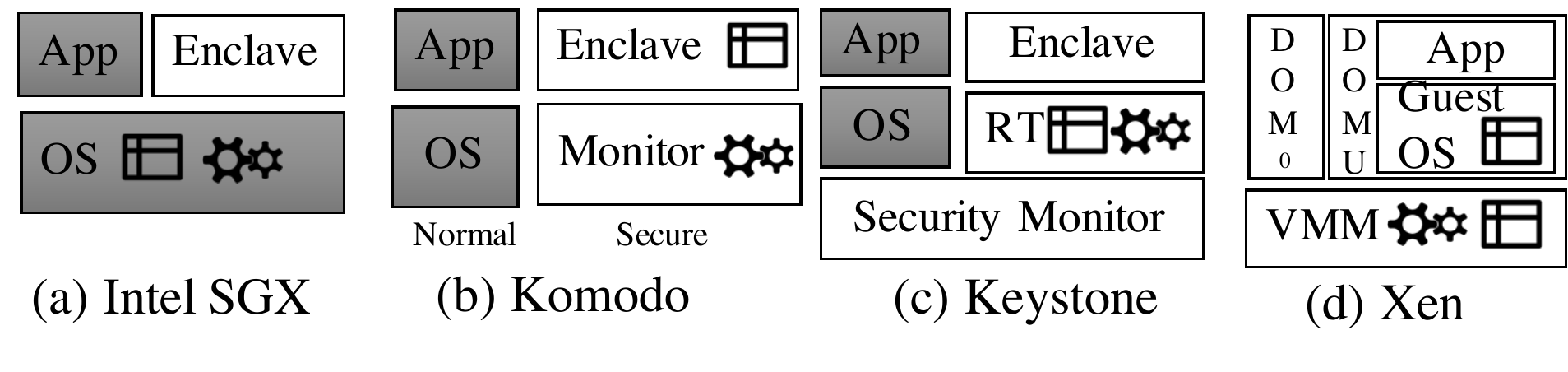}
\caption{
Memory Management Designs (the shaded area is untrusted).
(a) Intel SGX: the untrusted OS manages the memory and translates
virtual-to-physical address. (b) Komodo: the page-tables are inside
the enclave but the monitor creates the mappings. (c) \codename:
delegates page management to enclave with its own page table. (d) XEN:
the hypervisor performs the page management, there are two page tables
in this setup.
}
\label{fig:mmu_diff}
\end{figure}

\subsection{Post-creation In-enclave Page Management}
\codename has a different memory management design from the existing
TEEs (see Figure~\ref{fig:mmu_diff}). It uses the OS generated
page-tables for the initialization and then delegates the
virtual-to-physical memory mapping entirely to the enclave during
the execution. Since \riscv provides per-hardware-thread views of the
physical memory via the machine-mode and the PMP registers, it allows
\codename to have multiple concurrent and potentially multi-threaded
enclaves to access the disjoint physical memory partitions. With an
isolated S-mode inside the enclave, \codename can execute its own
virtual memory management which manipulates the enclave-specific
page-tables. Page tables are always inside the isolated enclave memory
space. By leaving the memory management to the enclave we: (a) support
several plugins for memory management which are not possible in any
of the existing TEEs (see Section~\ref{sec:mem_plugins}); (b) remove
controlled side-channel attacks as the host OS cannot modify or
observe the enclave virtual-to-physical mapping.
\section{\codename Framework}
\label{sec:framework}

\begin{figure}
\includegraphics[width=\columnwidth]{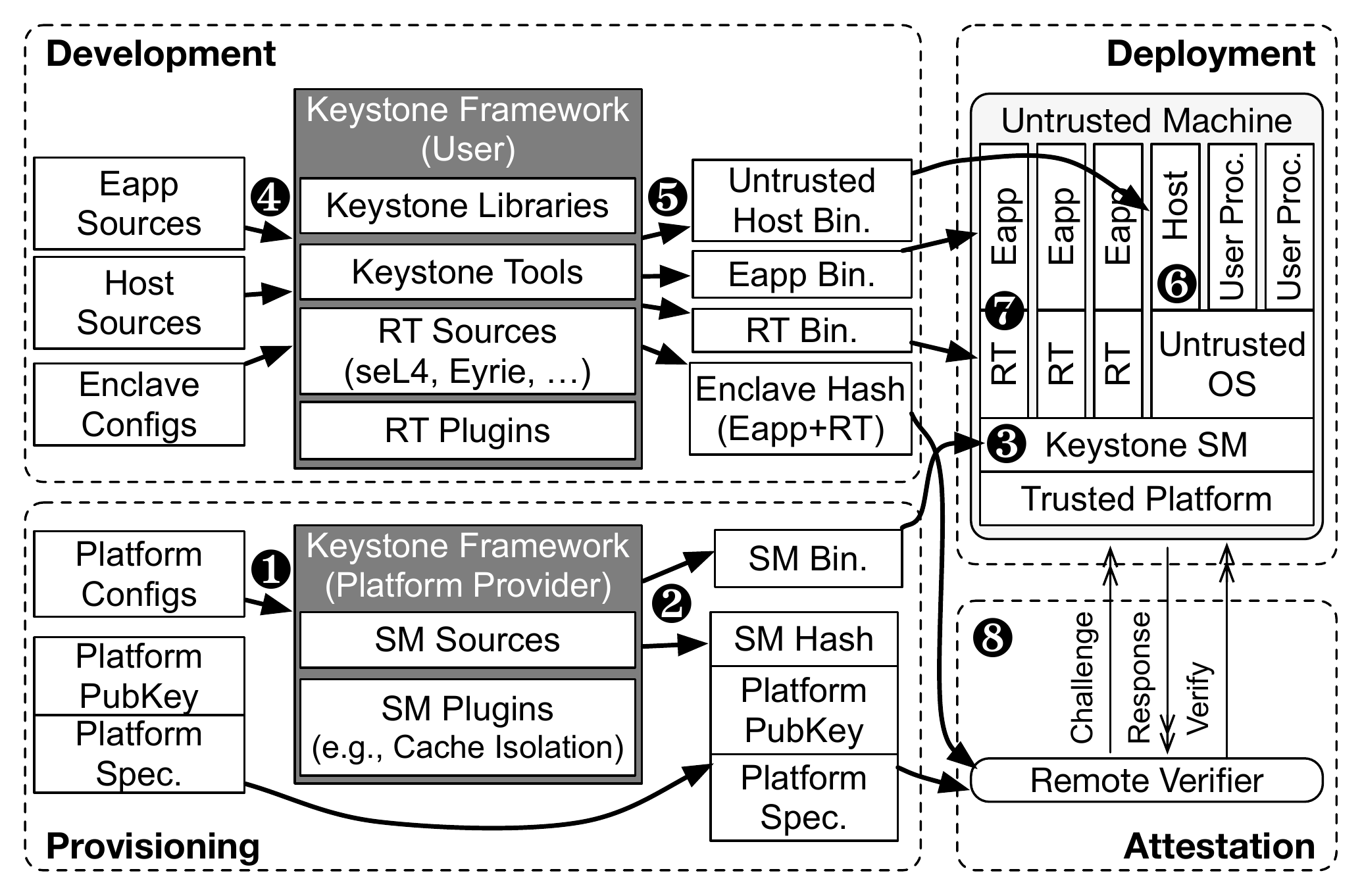}
\caption{ End-to-end Overview of \codename Framework. 
\ding{182} The platform provider configures the \sm. 
\ding{183} \codename compiles and generates the \sm boot image. 
\ding{184} The platform provider deploys the \sm. 
\ding{185} The developer writes an \eapp and configures the enclave. 
\ding{186} \codename builds the binaries and computes their 
measurements. 
\ding{187} The untrusted host binary is deployed to the machine. 
\ding{188} The host deploys the \rt and the \eapp and initiates the 
enclave creation.
\ding{189} A remote verifier can perform the attestation based on a 
known platform specifications, the keys, and the \sm/enclave measurements.}
\vspace{-12pt}
\label{fig:keystone_workflow}
\end{figure}

Figure~\ref{fig:keystone_workflow} details the steps from \codename
provisioning to \eapp deployment. Independently, the \eapp and the \rt
developers use \codename tools and libraries to write \eapps and
plugins. They can specify the plugins they want to use at the enclave
compile time. \sm plugins that expose additional hardware or security
functionality are determined at compile and provisioning time. We now
demonstrate our various \codename plugins.

\subsection{Enclave Memory Management Plugins}
\label{sec:mem_plugins}

Each enclave can run both S-mode and U-mode code. Thus, it has the
privileges to manage the enclave memory without crossing the
isolation boundaries. By default, \codename enclaves may occupy a
fixed contiguous physical memory allocated by the OS, with a
statically-mapped virtual address space at load time. Although this is
suitable for some embedded applications, it limits the memory usage of
most legacy applications. To this end, we describe several optional
plugins which enable flexible memory management of the enclave.

\paragraph{Free-memory.}
We built a plugin that allows the \eyrie \rt to perform page table
changes. It also lets the enclave reserve physical memory without
mapping it at creation time. This unmapped (hence, free) memory region
is not included in the enclave measurement and is zeroed before
beginning the \eapp execution. The free-memory plugin is required for
other more complex memory plugins.

\paragraph{Dynamic Resizing.}
Statically pre-defined maximum enclave size and subsequent static
physical / virtual memory pre-allocations prevent the enclave from 
scaling dynamically based on workload. It also complicates the process
of porting existing applications to \eapps. To this end, \codename
allows the \eyrie \rt to request dynamic changes to the physical
memory boundaries of the enclave. The \eyrie \rt may request that the
OS make an $\tt{extend}$ SBI call to add contiguous physical pages to
the enclave memory region. If the OS succeeds in such an allocation,
the \sm increases the size of the enclave and notifies the \eyrie \rt
who then uses the free memory plugin to manage the new physical pages.

\paragraph{In-Enclave Self Paging.}
We implemented a generic in-enclave page swapping plugin for the
\eyriert. It handles the enclave page-faults and uses a generic page
backing-store that manages the evicted page storage and retrieval. Our
plugin uses a simple random \eapp-only page eviction policy. It works
in conjunction with the free memory plugin for virtual memory
management in the \eyrie \rt. Put together, they help to alleviate the
tight memory restrictions an enclave may have due to the limited DRAM
or the on-chip memory size~\cite{eleos, varys, cosmix}.

\begin{figure}
\center
\includegraphics[width=\columnwidth]{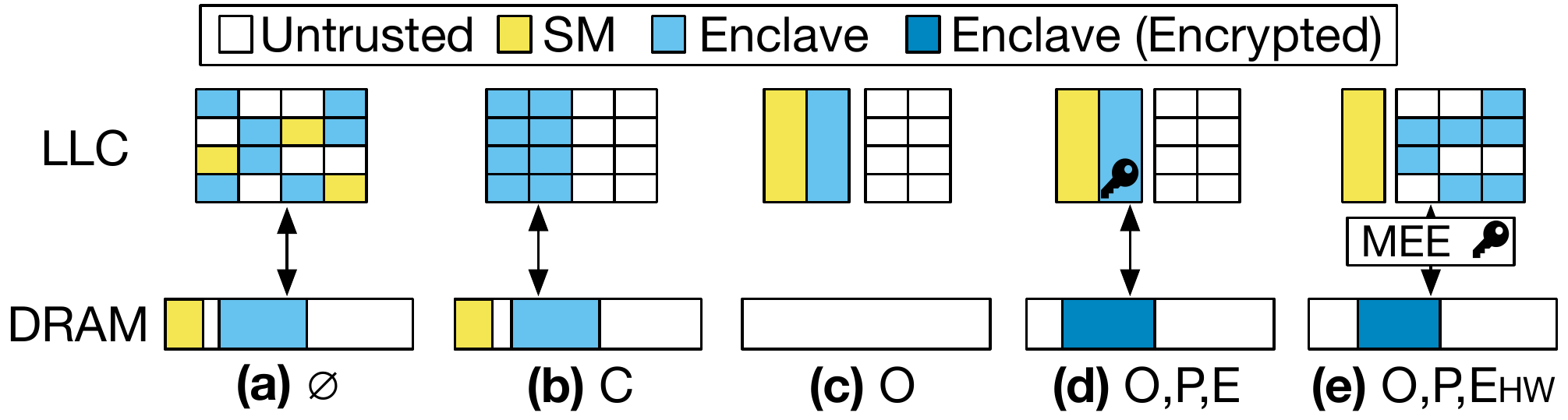}
\caption{\codename memory Model for the combination of various plugins.
 \textbf{$\varnothing$}: no plugin,
 \textbf{C}: cache partitioning, \textbf{O}: on-chip scratchpad,
 \textbf{P}: enclave self-paging, \textbf{E}: software memory 
 encryption
 \textbf{E$_{HW}$}: hardware memory encryption.
}
\label{fig:plugin_schematic}
\end{figure}

\paragraph{Protecting the Page Content Leaving the Enclave.}
When the enclave handles its own page fault, it may attempt to evict
the pages out of the secure physical memory (either an on-chip memory
or the protected portion of the DRAM). When these pages have to be
copied out, their content needs to be protected. Thus, as part of the
in-enclave page management, we implement a backing-store layer that
can include page encryption and integrity protection to allow for the
secure content to be paged out to the insecure storage (DRAM regions
or disk). The protection can be done either in the software as a part
of the \codename \rt (Figure~\ref{fig:plugin_schematic}(d)) or by a
dedicated trusted hardware unit---a memory encryption engine
(MEE)~\cite{general-purpose-mee}---with an \sm plugin
(Figure~\ref{fig:plugin_schematic}e). Admittedly, this incurs
significant design challenges in efficiently storing the metadata and
performance optimizations. \codename design is agnostic to the
specific integrity schemes and can reuse the existing
mechanisms~\cite{merkletree, bonsaitree}.

\subsection{SM Plugins}
\label{sec:security_plugins}
\codename can easily integrate with optional hardware to provide
stronger security guarantees at the cost of various trade-offs. We
demonstrate several SM plugins that can defend the enclave against a
physical attacker or cache side-channel attacks using the \hifivesoc
L2 memory controller. 

\paragraph{Secure On-chip Memory.}
To protect the enclaves against a physical attacker who has access to
the DRAM, we implemented an \emph{on-chip memory} plugin
(Figure~\ref{fig:plugin_schematic}(c)). It allows the enclave to
execute without the code or the data leaving the chip package. On the
\hifivesoc, we dynamically instantiate a scratchpad memory of up to
$2$MB via the L2 memory controller to generate a usable on-chip memory
region. An enclave requesting to run in the on-chip memory loads
nearly identically to the standard procedure with the following
changes: (a) the host loads the enclave to the OS allocated memory
region with modified initial page-tables referencing the final
scratchpad address; and (b) the \sm plugin copies the standard enclave
memory region into the new scratchpad region before the measurement.
Any context switch to the enclave now results in an execution in the
scratchpad memory. This plugin uses only our basic enclave life-cycle
hooks for the platform-specific plugins and does not require further
modification of the \sm. The only other change required was a
modification of the untrusted enclave loading process to make it aware
of the physical address region that the scratchpad occupies. No
modifications to the \eyriert or the \eapps are required.

\paragraph{Cache Partitioning.}
Enclaves are vulnerable to cache side-channel attacks from the
untrusted OS and the applications on the other cores via a shared
cache. To this end, we build a cache partitioning plugin using two
hardware capabilities: (a) the L2 cache controller's
\textit{waymasking} primitive similar to Intel's CAT~\cite{intelCAT};
(b) PMP to way-partition the L2 cache transparently to the OS and the
enclaves. Our plugin enforces effective {\em non-interference} between
the partitioned domains, see Figure~\ref{fig:plugin_schematic}(b).
Upon a context switch to the enclave, the cache lines in the partition
are flushed. During the enclave execution, only the cache lines from
the enclave physical memory are in the partition and are thus
protected by PMP. The adversary cannot insert cache lines in this
partition during the enclave execution due to the line replacement
way-masking mechanism. The net effect is that the adversary
(\attackerscc) gains no information about the evictions, the resident
lines, or the residency size of the enclave's cache. Ways are
partitioned at runtime and are available to the host whenever the
enclave is not executing even if paused.

\subsection{Functionality Plugins}
\label{sec:function_plugins}

\paragraph{Edge Call Interface.}
The \eapp cannot access the non-enclave memory in \codename. If it
needs to read/write the data outside the enclave, the \eyrie \rt
performs {\em edge calls} on its behalf. Our edge call, which is
functionally similar to RPC, consists of an index to a function
implemented in the untrusted host application and the parameters to be
passed to the function. \eyrie tunnels such a call safely to the
untrusted host, copies the return values of the function back to the
enclave, and sends them to the \eapp. The copying mechanism requires
\eyrie to have access to a buffer shared with the host. To enable
this: (a) the OS allocates a shared buffer in the host memory space
and passes the address to the \sm at enclave creation; (b) the \sm
passes the address to the enclave so the \rt may access this memory;
(c) the \sm uses a separate PMP entry to enable OS access to this
shared buffer. All the edge calls have to pass through the \eyriert as
the \eapp does not have access to the shared memory virtual mappings.
This plugin can be used to add support for syscalls, IPC,
enclave-enclave communication, and so on.

\paragraph{Proxied System Calls.}
We allow the proxying of some syscalls from the \eapp to the host
application by re-using the edge call interface. The user host
application then invokes the syscall on an untrusted OS on behalf of
the \eapp, collects the return values, and forwards them to the \eapp.
\codename can utilize existing defenses to prevent Iago
attacks~\cite{iago} via this interface~\cite{besfs, cosmix,
graphene-sgx}.

\paragraph{In-enclave System Calls.}
For appropriate syscalls (e.g., mmap, brk, getrandom), \codename
invokes an \sm interface or executes them in \eyrie to return the
results to the \eapp.

\paragraph{Multi-threading.}
We run multi-threaded \eapps by scheduling all the threads on the same
core. We do not support parallel multi-core enclave execution yet.

\subsection{Other \codename Primitives}
\codename supports following other standard TEE primitives.

\paragraph{Secure Boot.}
A \codename root-of-trust can be either a tamper-proof software (e.g.,
a zeroth-order bootloader) or hardware (e.g., crypto engine). At each
CPU reset, the root-of-trust (a) measures the \sm image, (b) generates
a fresh attestation key from a secure source of randomness, (c) stores
it to the \sm private memory, and (d) signs the measurement and the
public key with a hardware-visible secret. These are standard
operations, can be implemented in numerous
ways~\cite{sanctum-secure-boot,microsemi-secure-boot}. \codename does
not rely on a specific implementation. For completeness, currently,
\codename simulates secure boot via a modified first-stage bootloader
to performs all the above steps.

\paragraph{Secure Source of Randomness.}
\codename provides a secure \sm SBI call, $\tt{random}$, which
returns a $64$-bit random value. \codename uses a hardware source of
randomness if available or can use other well-known
options~\cite{mowery2013welcome} if applicable.

\paragraph{Trusted Timer.}
\codename allows the enclaves to access the read-only
hardware-maintained timer registers via the $\tt{rdcycle}$
instruction. The \sm supports standard timer SBI calls.

\paragraph{Remote Attestation.}
\codename \sm performs the measurement and the attestation based on
the provisioned key. Enclaves may request a signed attestation from
the \sm during runtime. \codename uses a standard scheme to bind the
attestation with a secure channel construction~\cite{komodo,
sanctum-secure-boot} by including limited arbitrary data (e.g.,
Diffie-Hellman key parameters) in the signed attestation report. Key
distribution~\cite{sgx-seal}, revocation~\cite{PKI-CRL}, attestation
services~\cite{intel-attestation-service}, and anonymous
attestation~\cite{direct-anonymous-attestation} are orthogonal
challenges.

\paragraph{Secure Delegation.}
\codename delegates the traps (i.e., interrupts and exceptions) raised
during the enclave execution to the \rt via the interrupt delegation
registers. The \rt invokes the appropriate handlers inside the enclave
for user-defined exceptions and may forward other traps to the
untrusted OS via the \sm.

\paragraph{Monotonic Counters \& Sealed Storage.}
Enclaves may need monotonic counters for protection against rollback
attacks and versioning~\cite{sgx-explained}. \codename can support
monotonic counters by keeping a limited counter state in the \sm
memory. \codename can support sealed storage~\cite{sgx-seal} in the
future.
\section{Security Analysis}
\label{sec:secanalysis}

We argue the security of the enclave, the OS, and the \sm based on the
threat model outlined in Section~\ref{sec:threat_model}.

\subsection{Protection of the Enclave}
\label{sec:enclave_sec}
\codename attestation ensures that any modification of the \sm, \rt,
and the \eapp is visible while creating the enclave. During the
enclave execution, any direct attempt by an \attackersw to access the 
enclave memory (cached or uncached) is defeated by PMP. All the
enclave data structures can only be updated in the enclave or the \sm,
both of them are isolated from direct access. Subtle attacks such as
controlled side channels (\attackerccs) are not possible in \codename
because enclaves have dedicated page management and in-enclave page
tables. This ensures that any enclave executing with any \codename
instantiated TEE is always protected against the above attacks.

\paragraph{Mapping Attacks.}
The \rt is trusted, does not intentionally create malicious virtual to
physical address mappings~\cite{inktag} and ensures that the mappings
are valid. The \rt initializes the page tables either during the
enclave creation or loads the pre-allocated (and \sm validated) static
mappings. During the enclave execution, the \rt ensures that the
layout is not corrupted while updating the mappings (e.g., via mmap).
When the enclave gets new empty pages, say via the dynamic memory
plugin, the \rt checks if they are safe to use before mapping them to
the enclave. Similarly, if the enclave is removing any pages, the \rt
scrubs their content before returning them to the OS.

\paragraph{Syscall Tampering Attacks.}
If the \eapp and the \rt invoke untrusted functions implemented in the
host process and/or execute the OS syscalls, they are susceptible to
Iago attacks and system call tampering attacks~\cite{iago,
ports-garfinkel}. \codename can re-use the existing shielding systems
~\cite{scone, besfs, graphene-sgx} as plugins to defend the enclave 
against these attacks.

\paragraph{Side-channel Attacks.}
\codename thwarts cache side-channel attacks
(Section~\ref{sec:security_plugins}). Enclaves do not share any state
with the host OS or the user application and hence are not exposed to
controlled channel attacks. The \sm performs a clean context switch
and flushes the enclave state (e.g., TLB). The enclave can defend
itself against explicit or implicit information leakage via the SM or
the edge call API with known defenses~\cite{msr-pldi16, moat}. Only
the \sm can see other enclave events (e.g., interrupts, faults), these
are not visible to the host OS. Timing attacks against the \eapp are
out of scope.

\subsection{Protecting the Host OS}
\label{sec:os_sec}
\codename \rts execute at the same privilege level as the host OS,
so an \attackersw is our case is stronger than in SGX. We ensure that
the host OS is not susceptible to new attacks from the enclave because
the enclave cannot: (a) reference any memory outside its allocated
region due to the \sm PMP-enforced isolation; (b) modify page tables
belonging to the host user-level application or the host OS; (c) 
pollute the host state because the \sm performs a complete context
switch (TLB, registers, L1-cache, etc.) when switching between an
enclave and the OS; (d) DoS a core as the \sm watchdog timer assures
that the enclave returns control to the OS.

\subsection{Protection of the \sm}
\label{sec:sm_sec}
The \sm naturally distrusts all the lower-privilege software
components (\eapp{}s, \rt{}s, host OS, etc.). It is protected from an
\attackersw because all the \sm memory is isolated using PMP and is
inaccessible to any enclave or the host OS. The \sm SBI is another
potential avenue of attack. \codename's \sm presents a narrow,
well-defined SBI to the S-mode code. It does not do complex resource
management and is small enough to be formally verified~\cite{komodo,
serval}. The \sm is only a reference monitor, it does not require
scheduled execution time, so an \attackerdos is not a concern. The \sm
can defend against an \attackerscc and an \attackersct with known
techniques~\cite{time-sc-survey, dawg}.

\subsection{Protection Against Physical Attackers}
\label{sec:phy_attacker}

\codename can protect against a physical adversary by a combination of
plugins and a proposed modification to the bootloader. Similar to
\cite{chen2008scratchpad}, we use a scratchpad to store the decrypted
code and data, while the supervisor mode component (\eyrie plugins)
handles the page-in and page-out.

\textbf{The enclave} itself is protected by the on-chip memory plugin
(in \sm) and the paging plugins (in \eyrie), with encryption and
integrity protection on the pages leaving the on-chip memory. The page
backing-store is a standard PMP protected physical memory region now
containing only the encrypted pages, similar in concept to the SGX
EPC. This fully guarantees the confidentiality and integrity of the
enclave code and data from an attacker with complete DRAM control. 

\textbf{The \sm} should be executed entirely from the on-chip memory.
The \sm is statically sized and has a relatively small in-memory
footprint ($<150$Kb). On the \hifivesoc, this would involve
repurposing a portion of the L2 loosely-integrated memory (LIM) via a
modified trusted bootloader.

With all of these plugins and techniques in place, the only content
present outside of the chip package is either untrusted (host, OS,
etc.) or is encrypted and integrity protected (e.g., swapped enclave
pages). \codename accomplishes this with no hardware modifications.
\section{Implementation}
\label{sec:implementation}
We implemented our \textbf{\sm} on top of the Berkeley Boot Loader
(bbl)~\cite{bbl-riscv-pk}. It supports M-mode trap handling, boot, and
other features. We simulated the primitives that require hardware
support (randomness, root-of-trust, etc.) and provided well-defined
interfaces. All the plugins in Section~\ref{sec:framework} are
available as compile-time options. Memory encryption is implemented
using software AES-128~\cite{aes-lib} and integrity protection is
partially implemented. We implemented the \textbf{\eyriert} from
scratch in C. We ported the \textbf{\sel microkernel}~\cite{sel4} to
\codename with $290$ \loc modifications to \sel for boot, memory
initialization, and interrupt handling. There are no inherent
restrictions to these two \rts, and we expect future development to
add further options. Our  user-land interface for interactions with
the enclaves is provided via a \textbf{Linux kernel module} that
creates a device endpoint (\texttt{/dev/\codename}).

We provide several libraries (edge-calls, host-side syscall endpoints,
attestation, etc.) in C and C++ for the host, the \eapp, and
interaction with the driver-provided Linux device. Our provided tools
generate the enclave measurements (hashes) without requiring \riscv
hardware, customize the \eyriert, and package the host application,
\eapp{}s, \rt into a single binary. We have a complete top-level build
solution to generate a bootable Linux image (based on the tooling for
the \hifivefull) for QEMU, FPGA softcores, and the \hifive containing
our \sm, the driver, and the enclave binaries. \codename is
open-source and available at \keystonegithub{}.

\paragraph{Writing \eapps.}
The \eapp developers can choose to port their legacy applications 
along with one of the default \rts provided by \codename. This
requires near-zero developer efforts. Alternatively, the developers
can use our SDK to write \eapp specifically for \codename. Lastly, 
\codename supports manually partitioning the applications such that
the security-sensitive parts of the logic execute inside the enclave,
while the rest of the code can execute in the untrusted host. Thus,
\codename framework gives the developers ample flexibility in
designing and implementing \eapps. 
\section{Evaluation}
\label{sec:evaluation}

We aim to answer the following questions in our evaluation:
\begin{enumerate}[label=(\bf{\em RQ\arabic*}), align=left, leftmargin=*] 

\item {\bf Modularity.}
Is the \codename framework viable in different configurations for real
applications?

\item {\bf TCB.}
What is the TCB of \codename-instantiated TEE in various deployment
modes?

\item {\bf Performance.}
How much overhead does \codename add to \eapp execution time?

\item {\bf Real-world Applications.}
Does \codename provide expressiveness with minimal developer efforts
for \eapps?

\end{enumerate} 

\paragraph{Experimental Setup.}
We used four different setups for our experiments; the \hifivefull
~\cite{HiFiveUnleashed} with a closed-source \hifivesoc (at 1GHz), and
three open-source \riscv processors: small Rocket (Rocket-S), default
Rocket~\cite{rocket}, and Berkeley Out-of-order Machine
(BOOM)~\cite{boom}. See Table~\ref{tab:hw_configs} for details. We
instantiated the open-source processors on FPGAs using
FireSim~\cite{firesim} which is responsible for emulating the cores at
1GHz. The host OS is buildroot Linux (kernel 4.15). All evaluation was
performed on the \hifive and the data is averaged over $10$ runs
unless otherwise specified.

\begin{table}[]
\footnotesize
\begin{tabular}{@{}lcccccccc@{}}
\toprule
\multirow{3}{*}{\textbf{Platform}} & \multicolumn{2}{c}{\multirow{2}{*}{\textbf{Core}}} & \multicolumn{2}{c}{\textbf{Cache Size}} & \multicolumn{2}{c}{\textbf{Latency}} & \multicolumn{2}{c}{\textbf{\# of TLB}} \\ 
 & \multicolumn{2}{c}{} & \multicolumn{2}{c}{\textbf{(KB)}} & \multicolumn{2}{c}{\textbf{(cycles)}} & \multicolumn{2}{c}{\textbf{Entries}} \\ \cmidrule(l){2-9} 
 & \textbf{\#} & \textbf{Type} & \textbf{L1-I/D} & \textbf{L2} & \textbf{L1} & \textbf{L2} & \textbf{L1} & \textbf{L2} \\ \midrule
\textbf{Rocket-S} & 1 & in-order & 8/8 & 512 & 2 & 24 & 8 & 128 \\ 
\textbf{Rocket} & 1 & in-order & 16/16 & 512 & 2 & 24 & 32 & 1024 \\ 
\textbf{BOOM} & 1 & OoO & 32/32 & 2048 & 4 & 24 & 32 & 1024 \\
\textbf{FU540} & 4 & in-order & 32/32 & 2048 & 2 & 12-15* & 32 & 128 \\ \bottomrule
\end{tabular}
\vspace{5pt}
\caption{
Hardware specification for each platform. L2 cache latency in FU540
(*) is based on estimation.
}
\vspace{-10pt}
\label{tab:hw_configs}
\end{table}

\subsection{Modularity \& Support}
\label{sec:modularity}



%

\begin{table}[]
\footnotesize
\centering
\begin{tabular}{lrr}
\toprule
\multicolumn{1}{c}{\textbf{Component}} & \multicolumn{1}{c}{\textbf{Runtime}} & \multicolumn{1}{c}{\textbf{\sm}} \\ \midrule
Base 			& 1800  & 1100 \\
Memory Isolation	& ---   &  500 \\
Free Memory         	&  300  & ---  \\
Dynamic Memory      	&  100  &   70 \\
Edge-call Handling  	&  300  &   30 \\
Syscalls            	&  450  & ---  \\
$\tt{libc}$ Environment &   50  & ---  \\
IO Syscall Proxying 	&  300  & ---  \\
Cache Partitioning  	& ---   &  300 \\
In-enclave Paging  	& 300   &  --- \\
On-chip Memory   	& ---   &  50  \\
\bottomrule
\end{tabular}
\vspace{10pt}
\caption{TCB Breakdowns (in \loc) for \eyrie \rt and \sm features.}
\vspace{-20pt}
\label{tab:runtime_tcb}
\end{table}

We outline the qualitative measurement of \codename flexibility in 
extending features, reducing TCB, and using the platform features.
Table~\ref{tab:runtime_tcb} shows the TCB breakdown of various
components (required and optional) for the \sm and \eyriert.

\paragraph{Extending \rts.} 
Most of the modifications (e.g., additional edge-call features)
require no changes to the \sm, and the \eappdev may enable them as
needed. Future additions (e.g., ports of interface shields) may be
implemented exclusively in the \rt. We also add support for a new \rt
by porting \sel to \codename and use it to execute various \eapps (See
Section~\ref{sec:case-studies}). \codename passes all the tests in 
\sel suite and incurs less than $1\%$ overhead on average over all
test cases.

\paragraph{Extending the \sm.}
The advantage of an easily modifiable \sm layer is noticeable when the
features require interaction with the core TEE primitives like memory
isolation. The \sm features were able to take advantage of the L2
cache controller on the FU540 to offer additional security protections
(cache-partitioning and on-chip isolation) without changes to the \rt
or \eapp.

\paragraph{TCB Breakdown.}
\codename comprises of the M-mode components (\texttt{bbl} and \sm),
the \rt, the untrusted host application, the \eapp, and the helper
libraries, of which only a fraction is in the TCB. The M-mode
component is $~10.7$ K\loc with a cryptographic library ($4$ K\loc),
pre-existing trap handling, boot, and utilities ($4.7$ K\loc), the \sm
($1.6$ K\loc), and \sm plugins ($400$ \loc).
A minimum \eyriert is $1.8$ K\loc, with plugins adding
further code as shown in Table~\ref{tab:runtime_tcb} up to a maximum
\eyriert TCB of $3.6$ K\loc.
The current maximum TCB for an \eapp running on our \sm and \eyriert
is thus a total of $~15$ K\loc.
TCB calculations were made using \texttt{cloc}~\cite{cloc} and
\texttt{unifdef}~\cite{unifdef}.

\subsection{Benchmarks}
\label{sec:benchmarks}

We use $4$ standard benchmark suites with a mix of CPU, memory, and
file I/O for system-wide analysis: \textbf{Beebs}, \textbf{CoreMark},
\textbf{RV8}, and \textbf{IOZone}. We report the overheads of the
cache partitioning plugin and physical attacker protection with RV8 as
an example of \codename trade-offs. In all the graphs, `other' refers
to the lifecycle costs for enclave creation, destruction, etc. 

\begin{figure}
\centering
\includegraphics[width=0.398\columnwidth]{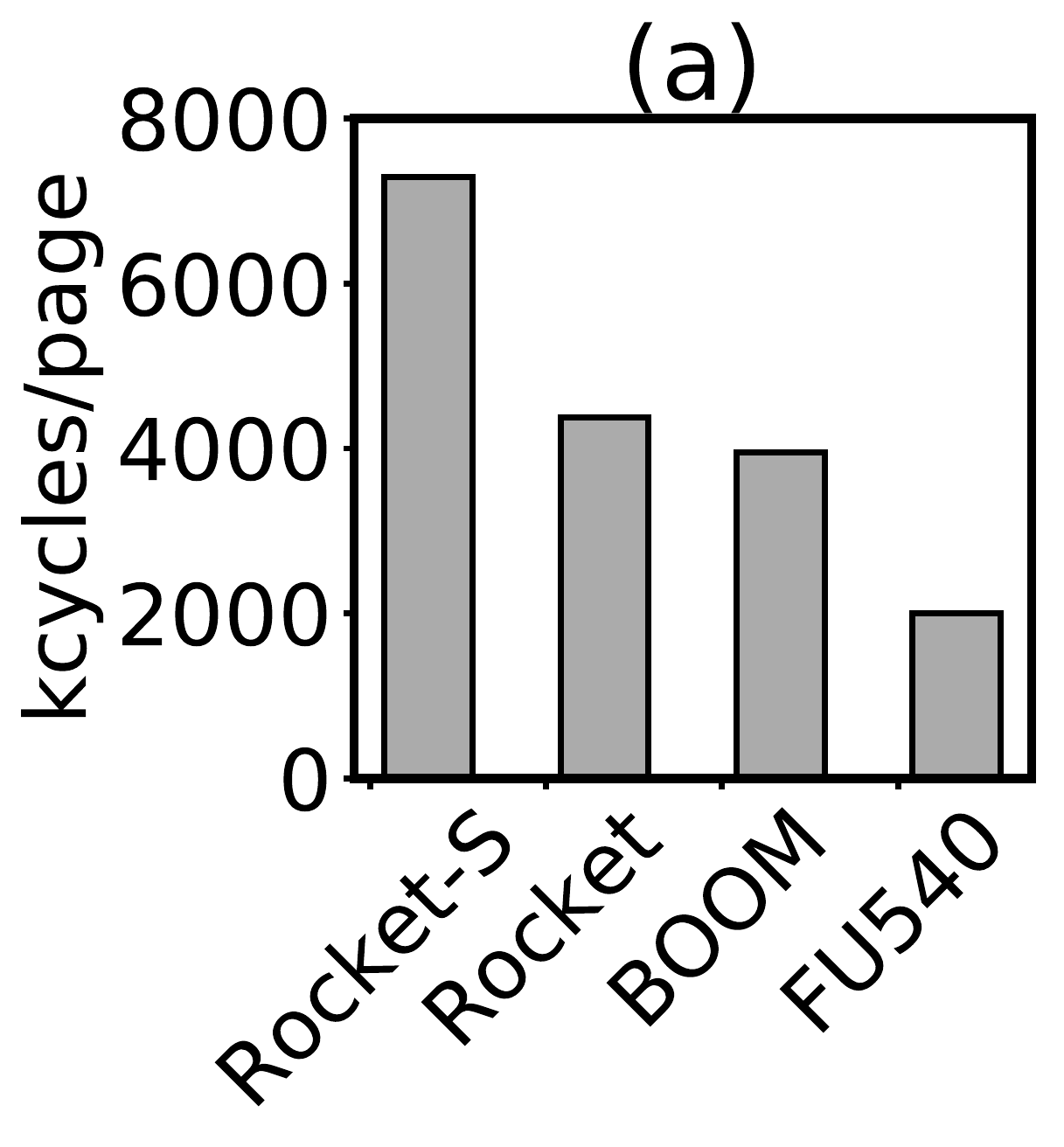}
\label{fig:breakdown_hash}
\includegraphics[width=0.592\columnwidth]{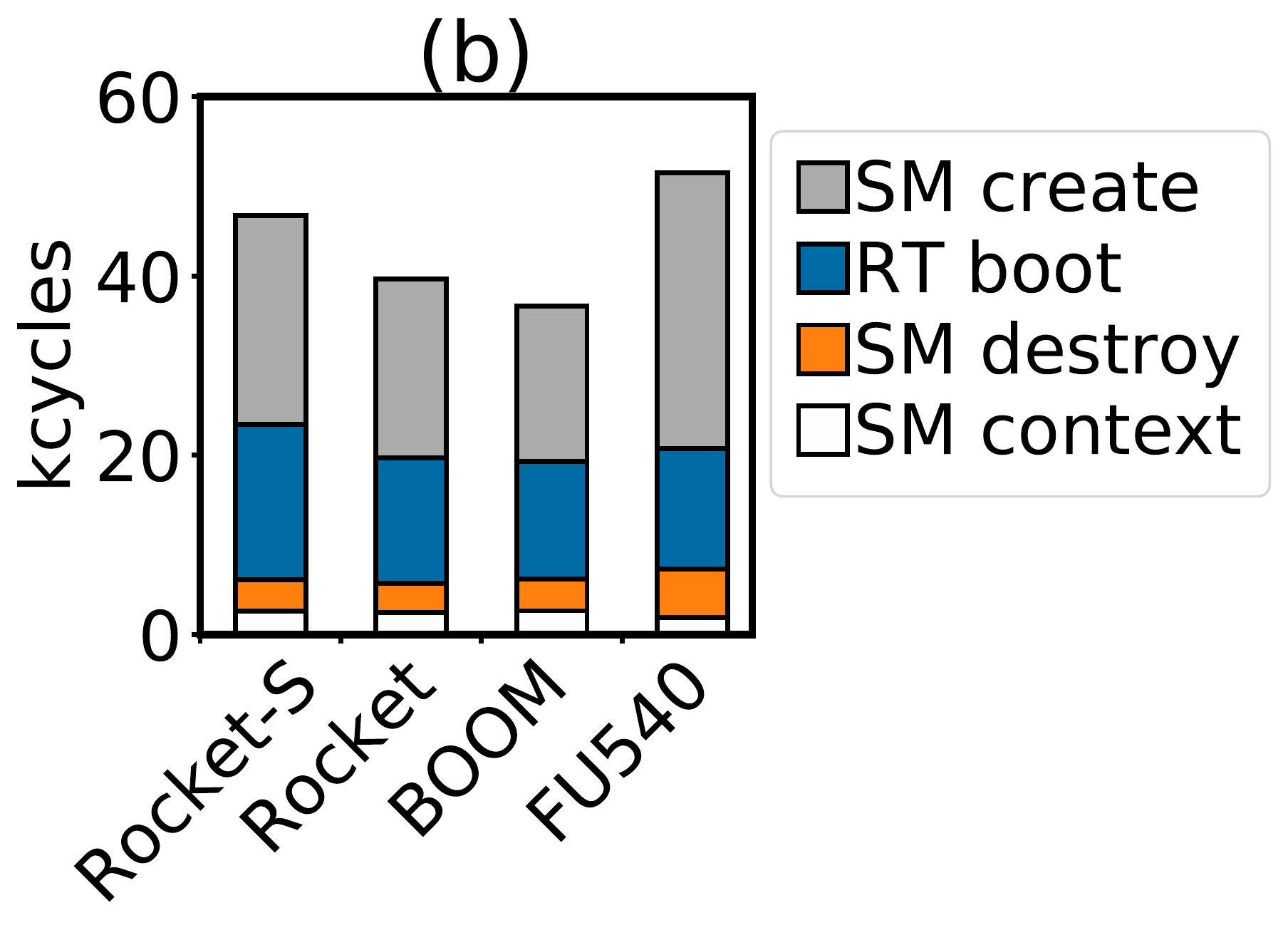}
\caption{
Breakdown of the operations during the enclave life-cycle. (a) shows
the enclave validation and the hashing duration, and (b) shows the
breakdown of other operations. (b) does not include the duration of
size-dependent operations such as the measurement in \texttt{create}
(Shown in (a)) and the memory cleaning in \texttt{destroy} (4K-11K
cycles/page).
}
\label{fig:breakdown}
\end{figure}

\paragraph{Common Operations.}
Figure~\ref{fig:breakdown} shows the breakdown of various enclave
operations. Initial validation and measurement dominate the startup
with $2$M and $7$M cycles/page for FU540 and Rocket-S due to an
unoptimized software implementation of SHA-3~\cite{sha3}. The
remaining enclave creation time totals $20$k-$30$k cycles. Similarly,
the attestation is dominated by the ed25519~\cite{ed25519} signing
software implementation with $0.7$M-$1.6$M cycles. These are both
one-time costs per-enclave and can be substantially optimized in
software or hardware. The most common \sm operation, context switches,
currently take between $1.8$K(FU540)-$2.6$K(Rocket-S) cycles depending
on the platform. Notably, creation and destruction of enclaves take 
longer on the FU540 ($4$-core), which can be attributed to the
multi-core PMP synchronization. 

\begin{figure}
\includegraphics[width=0.95\columnwidth]{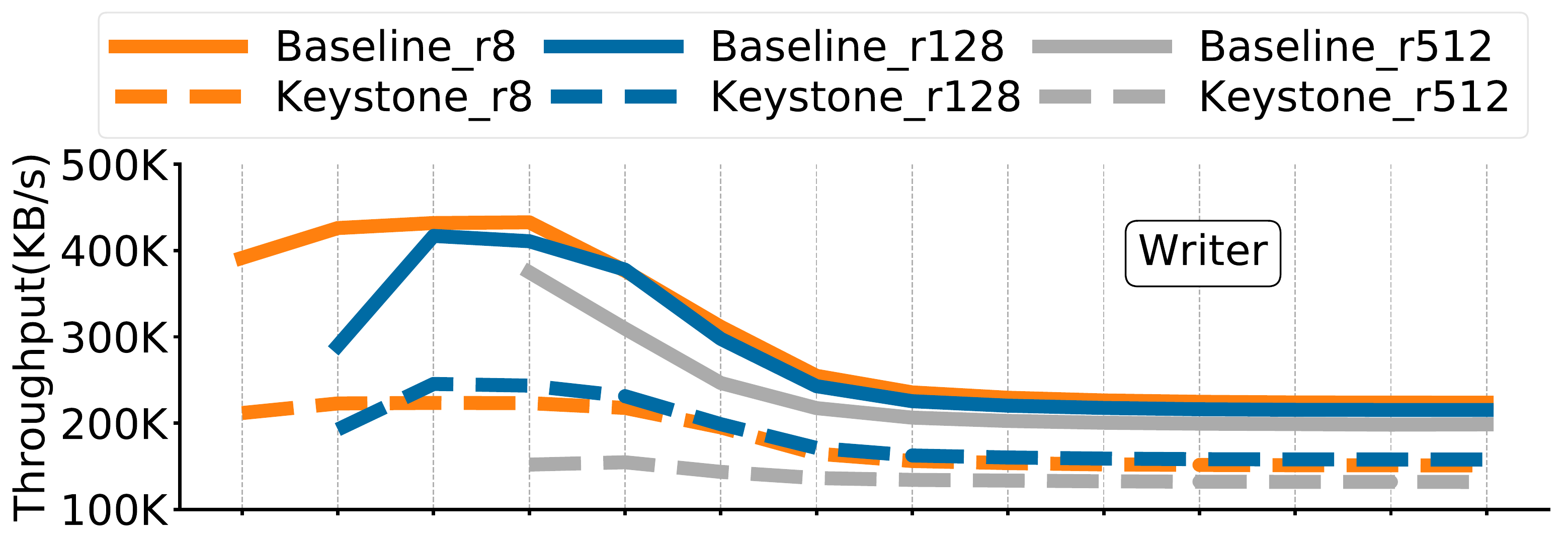}
\includegraphics[width=0.95\columnwidth]{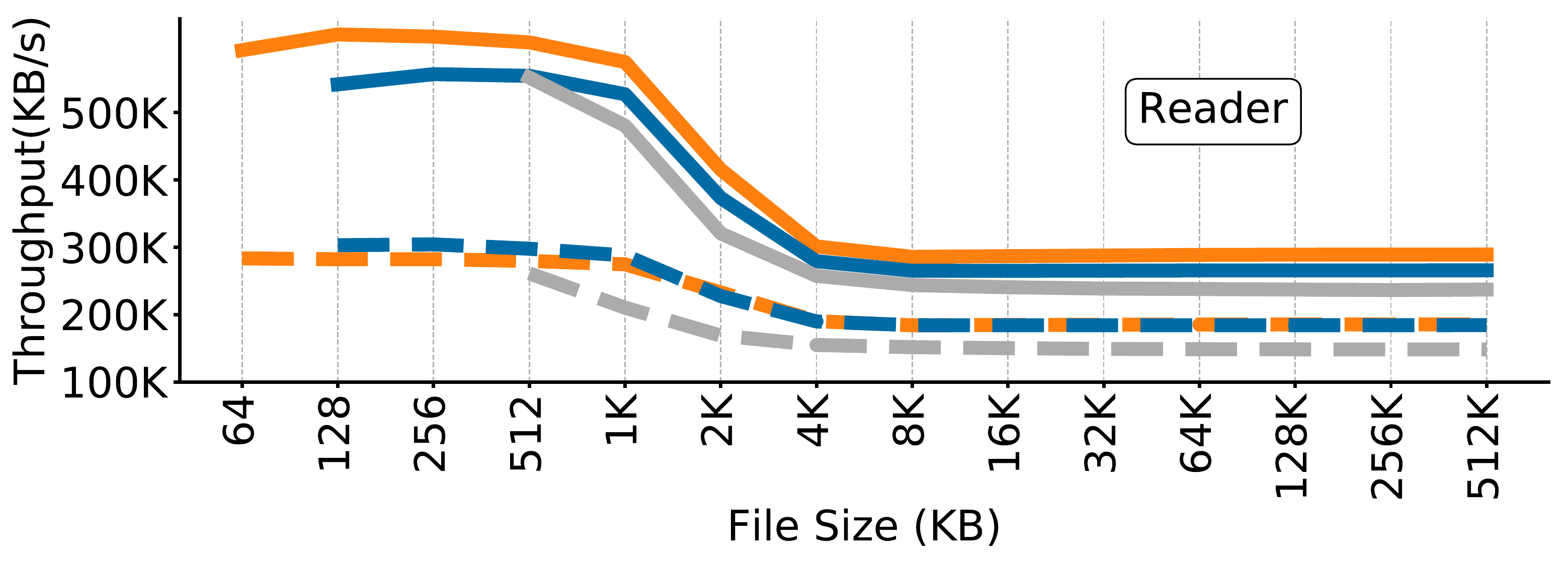}
\caption{
IOZone file operation throughput in \codename for various file and
record sizes (e.g., r8 represents 8KB record). We only show write and
read results due to limited space.
}
\label{fig:iozone}
\end{figure}

\paragraph{Standard Benchmarks Used as Unmodified \eapp Binaries.}
\textbf{ \em Beebs, CoreMark, and RV8.} As expected, \codename incurs
no meaningful overheads ($\pm 0.7\%$, excluding enclave
creation/destruction) for pure CPU and memory benchmarks.
\\
\textbf{\em IOZone.} All the target files are located on the untrusted
host and we tunnel the I/O system calls to the host application.
Figure~\ref{fig:iozone} shows the throughput plots of common
file-content access patterns. \codename experiences expected high
throughput loss for both write (avg. $36.2\%$) and read (avg.
$40.9.\%$). There are three factors contributing to the overhead: (a)
all the data crossing the privilege boundary is copied an additional
time via the untrusted buffer, (b) each call requires the \rt to go
through the edge call interface, incurring a constant overhead, and
(c) the untrusted buffer contends in the cache with the file buffers,
incurring an additional throughput loss on re-write (avg. $38.0\%$),
re-read (avg. $41.3\%$), and record re-write (avg. $55.1\%$)
operations. Since (b) is a fixed cost per system call, it increases
the overhead for the smaller record sizes.

\begin{figure}
\includegraphics[width=\columnwidth]{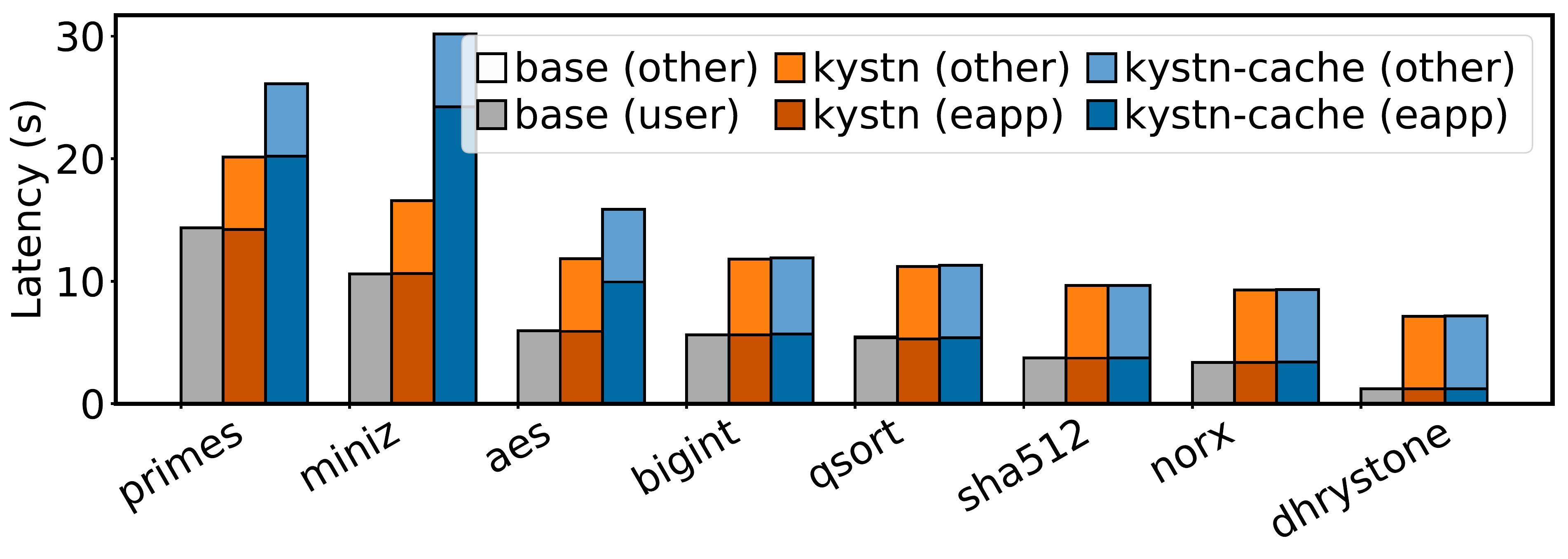}
\caption{
Total execution time comparison for the RV8 benchmarks. Each bar
consists of the duration of the application (\texttt{user} or
\texttt{eapp}), and the other overheads (\texttt{other}). \codename
(\texttt{\codename}) and \codename with cache partitioning plugin
(\texttt{\codename-cache}) are compared with the native execution in
Linux (\texttt{base}).
}
\label{fig:rv8}
\end{figure}

\paragraph{Cache Partitioning.}
The mix of pure-CPU and large working-set benchmarks in \textbf{RV8}
are ideal to demonstrate cache partitioning impact. For this
experiment, we grant $8$ of the $16$ ways in the FU540 L2 cache to the
enclave during execution (see Figure~\ref{fig:rv8}). Small working-set
tests show low overheads from cache flush on context switches, whereas
large working-set tests (primes, miniz, and aes) show up to $128.19\%$
overhead due to a smaller effective cache. The enclave initialization
latency is unaffected.

\begin{table}[]
\footnotesize
\centering
\resizebox{0.48\textwidth}{!}{%
\begin{tabular}{lrrrrl}
\toprule
\textbf{} & \multicolumn{4}{c}{\textbf{Overhead (\%)}} &\multirow{1}{*}{\textbf{\# of Page}} 
\\ \cline{1-5}  
\textbf{Plugins} & \textbf{$\varnothing$}     & \textbf{C}    & \textbf{O, P}   & \textbf{O, P, E} & \textbf{Faults}\\
\textbf{Adversary} & 
$\mathsf{A}_{SW,Cntrl}$\xspace & 
$\mathsf{A}_{SW,SC}$\xspace & 
$\mathsf{A}_{SW,SC}$\xspace &
$\mathsf{A}_{SW,SC,Phy_C}$\xspace & \textbf{(O, P)}  \\
\hline 
primes&   -0.9&   40.5&   65475.5&  * &   $66\times 10^6$ \\
miniz&   0.1&   128.5&   80.2&   615.5&   18341 \\
aes&   -1.1&   66.3&   1471.0&   4552.7&   59716 \\
bigint&   -0.1&   1.6&   0.4&   12.0&   168 \\
qsort&   -2.8&   -1.3&   12446.3&   26832.3&   285147 \\
sha512&   -0.1&   0.3&   -0.1&   -0.2&   0 \\
norx&   0.1&   0.9&   2590.1&   7966.4&   58834 \\
dhrystone&   -0.2&   0.3&   -0.2&   0.2&   0 \\
\bottomrule
\end{tabular}
}
\vspace{10pt}
\caption{Overhead of various plugins on RV8 benchmarks over native execution. 
\textbf{$\varnothing$}: no plugin, 
\textbf{C}: cache partitioning, 
\textbf{O}: on-chip scratch pad execution ($1$MB),
\textbf{P}: enclave self-paging,
\textbf{E}: software-based memory encryption.
*: does not complete in \textasciitilde$10$ hrs.
}
\label{tab:rv8_plugin_overhead}
\end{table}

\paragraph{Physical Attacker Protections.}
We ran the \textbf{RV8} suite with our on-chip execution plugins,
enclave self-paging, page encryption, and a DRAM backing page store 
(Table~\ref{tab:rv8_plugin_overhead}). A few \eapps (\texttt{sha512},
\texttt{dhrystone}), which fit in the $1$MB on-chip memory, incurs no
overhead and are protected even from \attackerphy. For the larger
working-set-size \eapps, the paging overhead increases depending on
the memory access pattern. For example, \texttt{primes} incurs the
largest amount of page faults because it allocates and randomly
accesses a $4MB$ buffer causing a page fault for almost every memory
access. Software-based memory encryption adds $2-4 \times$ more
overhead to page faults. These overheads can be alleviated by the
\codename framework if a larger on-chip memory or dedicated hardware
memory encryption engine is available as we discussed in
Section~\ref{sec:framework}.

\subsection{Case Studies}
\label{sec:case-studies}

We demonstrate how \codename can be adapted for a varied set of
devices, workloads, and application complexities with three
case-studies: (a) machine learning workloads for the client and
server-side usage, (b) a cryptographic library porting effort for
varied \rts, (c) a small secure computation application written
natively for \codename. The evaluation for these case-studies was 
performed on the \hifive board.

\paragraph{Porting Efforts \& Setup.}
We used the unmodified application code logic, hard-coded all the 
configurations and arguments for simplicity, and statically linked the
binaries against $\tt{glibc}$ or $\tt{musl}$~$\tt{libc}$ supported in
the \eyriert. We ported the widely used cryptographic library
\texttt{libsodium} to both \eyrie and \sel \rt trivially.

\begin{figure}
\includegraphics[width=\columnwidth]{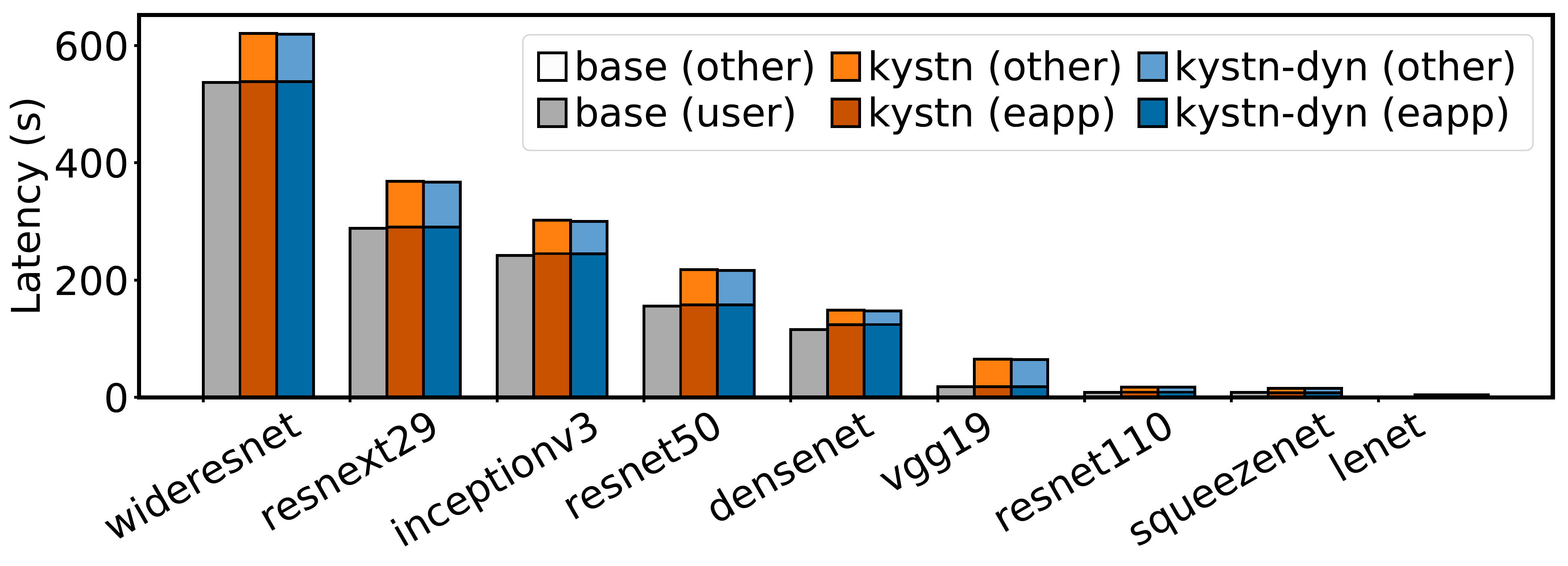}
\caption{ Comparison of inferencing duration for various Torch models
with and without dynamic resizing. Each bar consists of the duration
of the application (\texttt{user} or \texttt{eapp}), and the other
overheads (\texttt{other}). \codename (\texttt{\codename}) and
\codename with the dynamic resizing plugin (\texttt{\codename-dyn})
are compared with the native execution in Linux (\texttt{base}).
}
\label{fig:torch}
\end{figure}

\paragraph{Case-study 1: Secure ML Inference with Torch and \eyrie.}
We ran nine Torch-based models of increasing sizes with \eyrie on the
Imagenet dataset~\cite{imagenet} (see Table~\ref{tab:torch}). They
comprise $15.7$ and $15.4$K\loc of TH~\cite{torch-th} and
THNN~\cite{torch-thnn} libraries from Torch compiled with
$\tt{musl}$~$\tt{libc}$. Each model has an additional $230$ to $13.4$
K\loc of model-specific inference code~\cite{privado}. We performed
two sets of experiments: (a) execute the model inference code with
static maximum enclave size; (b) turn on the dynamic resizing plugin
so that the enclave extends its size on-demand when it executes.
Figure~\ref{fig:torch} shows the performance overheads for these two
configurations and non-enclaved execution baseline. 
\begin{table}[]
\footnotesize
\centering
\begin{tabular}{@{}lrrrrr@{}}
\toprule
\multicolumn{1}{c}{\textbf{Model}} & \multicolumn{1}{c}{\textbf{\begin{tabular}[c]{@{}c@{}}\# of \\ Layers\end{tabular}}} & \multicolumn{1}{c}{\textbf{\begin{tabular}[c]{@{}c@{}}\# of\\ Param\end{tabular}}}  & \multicolumn{1}{c}{\textbf{\begin{tabular}[c]{@{}c@{}}App\\ LOC\end{tabular}}} & \multicolumn{1}{c}{\textbf{\begin{tabular}[c]{@{}c@{}}Binary\\Size\end{tabular}}}  & \multicolumn{1}{c}{\textbf{\begin{tabular}[c]{@{}c@{}}Memory\\Usage\end{tabular}}}  \\ \midrule
Wideresnet   & 93   & 36.5M   & 1625  & 140MB  & 384MB  \\
Resnext29    & 102  & 34.5M   & 1910  & 123MB  & 394MB  \\
Inceptionv3  & 313  & 27.2M   & 5359  &  92MB  & 475MB  \\
Resnet50     & 176  & 25.6M   & 3094  &  98MB  & 424MB  \\
Densenet     & 910  &  8.1M   & 13399 &  32MB  & 570MB  \\
VGG19        & 55   & 20.0M   & 1088  &  77MB  & 165MB  \\
Resnet110    & 552  &  1.7M   & 9528  &   7MB  &  87MB  \\
Squeezenet   & 65   &  1.2M   & 914   &   5MB  &  52MB  \\
LeNet        & 12   &   62K   &	230   &  0.4MB &   2MB  \\
\bottomrule
\end{tabular}
\vspace{10pt}
\caption{Summary of the Torch Models. The model specification, 
workload characteristics, binary object size, and the total
enclave memory usage.}
\label{tab:torch}
\end{table}
\\
\textbf {\em Initialization Overhead.} is
noticeably high for both static size and dynamic resizing. It is
proportional to the \eapp binary size due to enclave page hashing.
Dynamic resizing reduces the initialization latency by $2.9$\% on
average as the \rt does not map free memory during enclave creation.
\\
\textbf{\em \eapp Execution Overhead.} We report an overhead between
$-3.12\%$(LeNet) to $7.35\%$(Densenet) for all the models with both
static enclave size and dynamic resizing. The cause for this is: (a)
\codename loads the entire binary in physical memory before it beings
\eapp execution, precluding any page faults for zero-fill-on-demand or
similar behavior, so smaller sized networks like LeNet execute faster
in \codename; (b) the overhead is primarily proportional to the number
of layers in the network, as more layers results in more memory
allocations and increase the number of $\tt{mmap}$ and $\tt{brk}$
syscalls. We used a small hand-coded test to verify that \eyrie \rt's
custom $\tt{mmap}$ is slower than the baseline kernel and incurs
overheads. So Densenet, which has the maximum number of layers
($910$), suffers from larger performance degradation. In summary, for
long-running \eapps, \codename incurs a fixed one-time startup cost
and the dynamic resizing plugin is indeed useful for larger \eapps.

\paragraph{Case-study 2: Secure ML with FANN and \sel.}
\codename can be used for small devices such as IoT sensors and
cameras to train models locally as well as flag events with model
inference. We ran FANN, a minimal (8K\loc C/C++) \eapp for embedded
devices with the \sel \rt to train and test a simple XOR network. The
end-to-end execution overhead is $0.36\%$ over running in \sel without
\codename.

\paragraph{Case-study 3: Secure Remote Computation.}
We designed and implemented a secure server \eapp (and remote client)
to count the number of words in a given message. We executed it with
\eyrie and no plugins. It performs attestation, uses libsodium to bind
a secure channel to the attestation report, then polls the host
application for encrypted messages using edge-calls, processes them
inside the enclave, and returns an encrypted reply to be sent to the
client. The \eapp consists of the secure channel code using libsodium
($60$ \loc), the edge-wrapping interface ($45$ \loc), and other logic
($60$ \loc). The host is $270$ \loc and the remote client is $280$
\loc. \codename takes $45$K cycles for a round-trip with an empty
message, secure channel, and message passing overheads. It takes
~$47$K cycles between the host getting a message and the enclave
notifying the host to reply.
\section{Related Work}
\label{sec:related}

\codename is the first framework for customizing TEEs.
Here, we survey existing TEEs and alternative approaches.

\paragraph{TEE Architectures \& Extensions.}
We summarize the design choices of the most recent TEEs and their
extensions in Table~\ref{tab:comparison}.  Of these, three major TEEs
are closely related to \codename: (a) Intel Software Guard Extension
(SGX) executes user-level code in an isolated virtual address space
backed by encrypted RAM pages~\cite{sgx}; (b) ARM TrustZone divides
the memory into two worlds (i.e., normal vs. secure) to run
applications in protected memory~\cite{trustzone}; and (c) Sanctum
uses the memory management unit (MMU) and cache partitioning to
isolate memory and prevent cache side-channel attacks~\cite{sanctum}. 
Several other TEEs explore design layers such as
hypervisors~\cite{hypersentry, hyperwall, trustvisor, cloudvisor,
nohype, overshadow, inktag, terra, sp, xoar, sp3, chaos}, physical
memory~\cite{sanctorum, mi6, bastion, flicker, oasis, trustlite,
aise}, virtual memory~\cite{timberv, iso-x, virtualghost, secureme,
sanctuary}, and process isolation~\cite{aegis, xom, xomos, drawbridge,
secureblue, ostia}. 

\paragraph{Re-purposing Existing TEEs for Modularity.}
One way to meet \codename design goals is to reuse the TEE solutions
available on commodity  CPUs. For each of these TEEs, it is possible
to enable a subset of  programming constructs (e.g., threading,
dynamic loading of binaries) by including a software management
component inside the enclave~\cite{optee, sierratee, t6, graphene-sgx,
haven}. On the other hand, all of the existing TEEs are hardware
extensions which are designed and implemented by the CPU manufacturer.
Thus, they do not allow users to access the programmable interface at
a layer underneath the untrusted OS. One way to simulate the
programmable layer is by adding a trusted hypervisor layer which then
executes an untrusted OS, but this approach inflates the TCB. Lastly,
none of these potential designs allow for adapting to threat models
and workloads.

\paragraph{Differences from Trusted Hypervisor}
\codename executes the enclave logic in the supervisor mode (\rt) and
the user mode (\eapp), while the machine mode code (\sm) merely checks
and enforces isolation boundaries. Although \codename may seem similar
to a trusted hypervisor, it does not implement or perform any resource
management, virtualization, or scheduling in the \sm. It merely checks
if the untrusted OS and the enclave (\rt, \eapp) are managing the
shared resources correctly. Thus, \codename \sm is more analogous to a
reference monitor~\cite{anderson1972computer, GrahamDenning1972}.

\paragraph{TEE Support.}
Several existing works  enhance existing TEE platforms. At the \sm
layer they optimize program-critical tasks~\cite{sanctum, timberv,
tz-rkp}, at \rt they target portability, functionality,
security~\cite{graphene-sgx, panoply, scone, haven, sancus, asylo, t6,
mesatee, optee, hyperapps, flexible-os, ms-openenclave}, and at \eapp
layer they reduce the developer efforts ~\cite{libseal, talos}.
Although these systems are a fixed configuration in the TEE design
space, they  provide valuable lessons for \codename feature design and
future optimization.

\paragraph{Enhancing the Security of TEEs.}
Better and secure TEE design has been a long-standing goal, with 
advocacy for security-by-design~\cite{bootstrap-trust,
seven-properties-highly-secure-devices}. We point out that \codename
is not vulnerable to a large class of side-channel
attacks~\cite{pte-sgx, cca-sgx, pigeonhole} by design, while
speculative execution attacks~\cite{spectre, foreshadow} are limited
to out-of-order \riscv cores (e.g., BOOM) and do not affect most SOC
implementations (e.g., Rocket). \codename can re-use known cache
side-channel defenses~\cite{mi6, dawg} as we demonstrated in
Section~\ref{sec:security_plugins}. Lastly, \codename can benefit from
various \riscv proposals underway to secure IO operations with
PMP~\cite{sifive-tee}. Thus, \codename either eliminates classes of
attacks or allows integration with existing techniques.

\paragraph{Formally Verified Hardware \& Software.}
TEE-like guarantees can be achieved orthogonally by a hardware and
software stack which is formally verified as resistant against all
classes of attacks that TEEs prevent. A careful and ground-up design
with verified components ~\cite{cheriabi,cheri, kami, sel4, certikos,
vale, vale-popl, certicoq, compcert-paper,evercrypt, everest-snapl,
verve, ironclad, deepspec, device-veri, hyperkernel, cogent} may
provide stronger guarantees, and \codename can help to explore
designs which combine these with hardware protection~\cite{komodo,
tap}.
\section{Conclusion}
\label{sec:conclusion}
We present \codename, the first framework for \programmable TEEs. With
our modular design, we showcase the use of \codename for several
standard benchmarks and applications on illustrative \rts and various
deployments platforms. \codename can serve as a framework for
research and prototyping better TEE designs.
\begin{acks}
We thank Srini Devdas and Ilia Lebedev for sharing the Sanctum
codebase and initial discussions. We thank Alexander Thomas and
Stephan Kaminsky for their help in building \codename and Jerry Zhao
for help in running \codename on BOOM. Thanks to the members of UCB
BAR for their help with the \hifivefull and FireSim setup. We thank
Paul Kocher and Howard Wu for their feedback on the early versions of
this draft. Thanks to the Privado team at Microsoft Research for 
sharing the torch code for our case-study. This material is in part
based upon work supported by the National Science Foundation under
Grant No. TWC-1518899 and DARPA N66001-15-C-4066. Any opinions,
findings, and conclusions or recommendations expressed in this
material are those of the author(s) and do not necessarily reflect the
views of the National Science Foundation. Research partially funded by
RISE Lab sponsor Amazon Web Services, ADEPT Lab industrial sponsors
and affiliates Intel, HP, Huawei, NVIDIA, and SK Hynix. Any opinions,
findings, conclusions, or recommendations in this paper are solely
those of the authors and do not necessarily reflect the position or
the policy of the sponsors.
\end{acks}
\balance
\def\UrlBreaks{\do\/\do-}
\bibliographystyle{plain}
\bibliography{keystone_v2}

\begin{thebibliography}{100}

\bibitem{aes-lib}
{Basic implementations of standard cryptography algorithms}.
\newblock \url{https://github.com/B-Con/crypto-algorithms}.

\bibitem{cloc}
{cloc - count lines of code}.
\newblock \url{https://github.com/AlDanial/cloc}.

\bibitem{HiFiveUnleashed}
{HiFive Unleashed - SiFive}.
\newblock \url{https://www.sifive.com/boards/hifive-unleashed}.

\bibitem{mesatee}
{MesaTEE: A Trustworthy Memory-safe Distributed Computing Tramework}.
\newblock \url{https://mesatee.org/}.

\bibitem{multizone}
{MultiZone Hex Five Security – The 1st Trusted Execution Environment for
  RISC-V}.
\newblock \url{https://hex-five.com/}.

\bibitem{ms-openenclave}
Open enclave sdk.
\newblock \url{https://openenclave.io/sdk/}.

\bibitem{optee}
{Open Portable Trusted Execution Environment - OP-TEE}.
\newblock \url{https://www.op-tee.org/}.

\bibitem{ed25519}
{Portable C Implementation of Ed25519, A high-speed high-security public-key
  signature system.}
\newblock \url{https://github.com/mit-sanctum/ed25519}.

\bibitem{poet}
Proof of elapsed time (poet) 1.0 specification — sawtooth v1.0.5
  documentation.
\newblock
  \url{https://sawtooth.hyperledger.org/docs/core/releases/1.0/architecture/poet.html}.

\bibitem{bbl-riscv-pk}
{RISC-V Proxy Kernel and Boot Loader}.
\newblock \url{https://github.com/riscv/riscv-pk}.

\bibitem{sierratee}
{SierraTEE Virtualization for ARM TrustZone and MIPS}.
\newblock \url{https://www.sierraware.com/open-source-ARM-TrustZone.html}.

\bibitem{anon-code-release}
{\codename Source Code Release}.
\newblock \keystoneurl{}.

\bibitem{torch-th}
{Standalone C TH library. C/CPU implementation of Tensors}.
\newblock \url{https://github.com/torch/TH}.

\bibitem{t6}
{T6 - Secure OS and TEE}.
\newblock \url{https://www.trustkernel.com/en/products/}.

\bibitem{sha3}
{Tiny SHA3}.
\newblock \url{https://github.com/mjosaarinen/tiny_sha3/}.

\bibitem{torch-thnn}
Torch/nn gathers nn's c implementations of neural network modules.
\newblock \url{https://github.com/torch/nn/tree/master/lib/THNN}.

\bibitem{unifdef}
{unifdef}.
\newblock \url{http://dotat.at/prog/unifdef/}.

\bibitem{cogent}
Sidney Amani, Alex Hixon, Zilin Chen, Christine Rizkallah, Peter Chubb, Liam
  O'Connor, Joel Beeren, Yutaka Nagashima, Japheth Lim, Thomas Sewell, Joseph
  Tuong, Gabriele Keller, Toby Murray, Gerwin Klein, and Gernot Heiser.
\newblock {Cogent}: Verifying high-assurance file system implementations.
\newblock ISCA'16.

\bibitem{certicoq}
Abhishek Anand, Andrew Appel, Greg Morrisett, Zoe Paraskevopoulou, Randy
  Pollack, Olivier~Savary Belanger, Matthieu Sozeau, and Matthew Weaver.
\newblock Certicoq: A verified compiler for coq.
\newblock CoqPL'17.

\bibitem{sgx-seal}
Ittai Anati, Shay Gueron, Simon~P Johnson, and Vincent~R Scarlata.
\newblock {Innovative Technology for CPU Based Attestation and Sealing}.
\newblock In {\em HASP}, 2013.

\bibitem{anderson1972computer}
James~P Anderson.
\newblock Computer security technology planning study.
\newblock Technical report, Anderson (James P) and Co Fort Washington PA, 1972.

\bibitem{riscv-priv-110}
Krste~Asanovi\'{c} Andrew~Waterman.
\newblock {The RISC-V Instruction Set Manual Volume II: Privileged
  Architecture}.
\newblock
  \url{https://content.riscv.org/wp-content/uploads/2017/05/riscv-privileged-v1.10.pdf},
  May 2017.

\bibitem{deepspec}
Andrew~W. Appel, Lennart Beringer, Adam Chlipala, Benjamin~C. Pierce, Zhong
  Shao, Stephanie Weirich, and Steve Zdancewic.
\newblock Position paper: the science of deep specification.
\newblock {\em Philosophical Transactions of the Royal Society of London A:
  Mathematical, Physical and Engineering Sciences}, 375(2104), 2017.

\bibitem{trustzone}
ARM.
\newblock {ARM Security Technology - Building a Secure System using TrustZone
  Technology. ARM Technical White Paper}.
\newblock 2013.

\bibitem{scone}
Sergei Arnautov, Bohdan Trach, Franz Gregor, Thomas Knauth, Andre Martin,
  Christian Priebe, Joshua Lind, Divya Muthukumaran, Daniel
  O{\textquoteright}Keeffe, Mark~L Stillwell, David Goltzsche, Dave Eyers,
  R{\"u}diger Kapitza, Peter Pietzuch, and Christof Fetzer.
\newblock {SCONE: Secure Linux Containers with Intel SGX}.
\newblock In {\em OSDI}, 2016.

\bibitem{rocket}
Krste Asanovi{\'c}, Rimas Avizienis, Jonathan Bachrach, Scott Beamer, David
  Biancolin, Christopher Celio, Henry Cook, Daniel Dabbelt, John Hauser, Adam
  Izraelevitz, Sagar Karandikar, Ben Keller, Donggyu Kim, John Koenig, Yunsup
  Lee, Eric Love, Martin Maas, Albert Magyar, Howard Mao, Miquel Moreto, Albert
  Ou, David~A. Patterson, Brian Richards, Colin Schmidt, Stephen Twigg, Huy Vo,
  and Andrew Waterman.
\newblock The rocket chip generator.
\newblock (UCB/EECS-2016-17), Apr 2016.

\bibitem{libseal}
Pierre-Louis Aublin, Florian Kelbert, Dan O'Keeffe, Divya Muthukumaran,
  Christian Priebe, Joshua Lind, Robert Krahn, Christof Fetzer, David Eyers,
  and Peter Pietzuch.
\newblock Libseal: Revealing service integrity violations using trusted
  execution.
\newblock In {\em EuroSys}, 2018.

\bibitem{talos}
Pierre-Louis Aublin, Florian Kelbert, Dan O'Keeffe, Divya Muthukumaran,
  Christian Priebe, Joshua Lind, Robert Krahn, Christof Fetzer, David~M. Eyers,
  and Peter~R. Pietzuch.
\newblock Talos : Secure and transparent tls termination inside sgx enclaves.
\newblock 2017.

\bibitem{tz-rkp}
Ahmed~M. Azab, Peng Ning, Jitesh Shah, Quan Chen, Rohan Bhutkar, Guruprasad
  Ganesh, Jia Ma, and Wenbo Shen.
\newblock Hypervision across worlds: Real-time kernel protection from the arm
  trustzone secure world.
\newblock In {\em CCS}, 2014.

\bibitem{hypersentry}
Ahmed~M. Azab, Peng Ning, Zhi Wang, Xuxian Jiang, Xiaolan Zhang, and Nathan~C.
  Skalsky.
\newblock Hypersentry: Enabling stealthy in-context measurement of hypervisor
  integrity.
\newblock In {\em CCS}, 2010.

\bibitem{haven}
Andrew Baumann, Marcus Peinado, and Galen Hunt.
\newblock {Shielding Applications from an Untrusted Cloud with Haven}.
\newblock In {\em OSDI}, 2014.

\bibitem{tesseract}
Iddo Bentov, Yan Ji, Fan Zhang, Yunqi Li, Xueyuan Zhao, Lorenz Breidenbach,
  Philip Daian, and Ari Juels.
\newblock Tesseract: Real-time cryptocurrency exchange using trusted hardware.
\newblock ePrint Archive, Report 2017/1153, 2017.

\bibitem{everest-snapl}
Karthikeyan Bhargavan, Barry Bond, Antoine Delignat-Lavaud, Cédric Fournet,
  Chris Hawblitzel, Catalin Hritcu, Samin Ishtiaq, Markulf Kohlweiss, Rustan
  Leino, Jay Lorch, Kenji Maillard, Jinyang Pang, Bryan Parno, Jonathan
  Protzenko, Tahina Ramananandro, Ashay Rane, Aseem Rastogi, Nikhil Swamy,
  Laure Thompson, Peng Wang, Santiago Zanella-Beguelin, and Jean-Karim
  Zinzindohoué.
\newblock Everest: Towards a verified, drop-in replacement of {HTTPS}.
\newblock In {\em Summit on Advances in Programming Languages (SNAPL)}, 2017.

\bibitem{vale}
Barry Bond, Chris Hawblitzel, Manos Kapritsos, K.~Rustan~M. Leino, Jacob~R.
  Lorch, Bryan Parno, Ashay Rane, Srinath Setty, and Laure Thompson.
\newblock Vale: Verifying high-performance cryptographic assembly code.
\newblock In {\em USENIX Security}, August 2017.

\bibitem{mi6}
Thomas Bourgeat, Ilia~A. Lebedev, Andrew Wright, Sizhuo Zhang, Arvind, and
  Srinivas Devadas.
\newblock {MI6:} secure enclaves in a speculative out-of-order processor.
\newblock 2018.

\bibitem{sanctuary}
Ferdinand Brasser, David Gens, Patrick Jauernig, Ahmad-Reza Sadeghi, and
  Emmanuel Stapf.
\newblock Sanctuary: Arming trustzone with user-space enclaves.
\newblock In {\em NDSS}, 2019.

\bibitem{securekeeper}
Stefan Brenner, Colin Wulf, David Goltzsche, Nico Weichbrodt, Matthias Lorenz,
  Christof Fetzer, Peter Pietzuch, and R\"{u}diger Kapitza.
\newblock Securekeeper: Confidential zookeeper using intel sgx.
\newblock In {\em Middleware}, 2016.

\bibitem{direct-anonymous-attestation}
Ernie Brickell, Jan Camenisch, and Liqun Chen.
\newblock Direct anonymous attestation.
\newblock In {\em CCS}, 2004.

\bibitem{foreshadow}
Jo~Van Bulck, Marina Minkin, Ofir Weisse, Daniel Genkin, Baris Kasikci, Frank
  Piessens, Mark Silberstein, Thomas~F. Wenisch, Yuval Yarom, and Raoul
  Strackx.
\newblock Foreshadow: Extracting the keys to the intel {SGX} kingdom with
  transient out-of-order execution.
\newblock In {\em {USENIX} Security}, 2018.

\bibitem{pte-sgx}
Jo~Van Bulck, Nico Weichbrodt, R{\"u}diger Kapitza, Frank Piessens, and Raoul
  Strackx.
\newblock Telling your secrets without page faults: Stealthy page table-based
  attacks on enclaved execution.
\newblock In {\em {USENIX}}, 2017.

\bibitem{boom}
Christopher Celio, David~A. Patterson, and Krste Asanovi{\'c}.
\newblock The berkeley out-of-order machine (boom): An industry-competitive,
  synthesizable, parameterized risc-v processor.
\newblock Technical Report UCB/EECS-2015-167, Jun 2015.

\bibitem{bastion}
D.~Champagne and R.~B. Lee.
\newblock Scalable architectural support for trusted software.
\newblock In {\em HPCA}, 2010.

\bibitem{graphene-sgx}
Chia che Tsai, Donald~E. Porter, and Mona Vij.
\newblock Graphene-sgx: A practical library {OS} for unmodified applications on
  {SGX}.
\newblock In {\em {USENIX ATC}}, 2017.

\bibitem{iago}
Stephen Checkoway and Hovav Shacham.
\newblock {Iago attacks: Why the System Call API is a Bad Untrusted RPC
  Interface}.
\newblock In {\em ASPLOS}, 2013.

\bibitem{chaos}
Haibo Chen, Fengzhe Zhang, Cheng Chen, Ziye Yang, Rong Chen, Binyu Zang, and
  Wenbo Mao.
\newblock {Tamper-Resistant Execution in an Untrusted Operating System Using A
  Virtual Machine Monitor}, 2007.

\bibitem{device-veri}
Hao Chen, Xiongnan~(Newman) Wu, Zhong Shao, Joshua Lockerman, and Ronghui Gu.
\newblock Toward compositional verification of interruptible os kernels and
  device drivers.
\newblock PLDI '16.

\bibitem{chen2008scratchpad}
Xi~Chen, Robert~P Dick, and Alok Choudhary.
\newblock Operating system controlled processor-memory bus encryption.
\newblock In {\em 2008 Design, Automation and Test in Europe}, pages
  1154--1159. IEEE, 2008.

\bibitem{overshadow}
Xiaoxin Chen, Tal Garfinkel, E.~Christopher Lewis, Pratap Subrahmanyam, Carl~A.
  Waldspurger, Dan Boneh, Jeffrey Dwoskin, and Dan~R.K. Ports.
\newblock {Overshadow: A Virtualization-Based Approach to Retrofitting
  Protection in Commodity Operating Systems}.
\newblock In {\em ASPLOS}, 2008.

\bibitem{ekiden}
R.~{Cheng}, F.~{Zhang}, J.~{Kos}, W.~{He}, N.~{Hynes}, N.~{Johnson},
  A.~{Juels}, A.~{Miller}, and D.~{Song}.
\newblock {Ekiden: A Platform for Confidentiality-Preserving, Trustworthy, and
  Performant Smart Contract Execution}.
\newblock {\em ArXiv e-prints}, 2018.

\bibitem{secureme}
Siddhartha Chhabra, Brian Rogers, Yan Solihin, and Milos Prvulovic.
\newblock {SecureME: A Hardware-software Approach to Full System Security}.
\newblock In {\em ICS}, 2011.

\bibitem{kami}
Joonwon Choi, Muralidaran Vijayaraghavan, Benjamin Sherman, Adam Chlipala, and
  Arvind.
\newblock Kami: A platform for high-level parametric hardware specification and
  its modular verification.
\newblock {\em Proc. ACM Program. Lang.}, 1(ICFP), 2017.

\bibitem{xoar}
Patrick Colp, Mihir Nanavati, Jun Zhu, William Aiello, George Coker, Tim
  Deegan, Peter Loscocco, and Andrew Warfield.
\newblock Breaking up is hard to do: Security and functionality in a commodity
  hypervisor.
\newblock In {\em SOSP}, 2011.

\bibitem{sgx-explained}
Victor Costan and Srinivas Devadas.
\newblock Intel sgx explained.
\newblock Cryptology ePrint Archive, Report 2016/086, 2016.
\newblock \url{http://eprint.iacr.org/2016/086}.

\bibitem{sanctum}
Victor Costan, Ilia Lebedev, and Srinivas Devadas.
\newblock Sanctum: Minimal hardware extensions for strong software isolation.
\newblock In {\em USENIX Security}, 2016.

\bibitem{virtualghost}
John Criswell, Nathan Dautenhahn, and Vikram Adve.
\newblock Virtual ghost: Protecting applications from hostile operating
  systems.
\newblock In {\em ASPLOS}, 2014.

\bibitem{xomos}
Mark~Horowitz David~Lie, Chandramohan A.~Thekkath.
\newblock {Implementing an Untrusted Operating System on Trusted Hardware}.
\newblock In {\em SOSP}, 2003.

\bibitem{cheriabi}
Brooks Davis, Robert N.~M. Watson, Alexander Richardson, Peter~G. Neumann,
  Simon~W. Moore, John Baldwin, David Chisnall, James Clarke, Nathaniel~Wesley
  Filardo, Khilan Gudka, Alexandre Joannou, Ben Laurie, A.~Theodore Markettos,
  J.~Edward Maste, Alfredo Mazzinghi, Edward~Tomasz Napierala, Robert~M.
  Norton, Michael Roe, Peter Sewell, Stacey Son, and Jonathan Woodruff.
\newblock Cheriabi: Enforcing valid pointer provenance and minimizing pointer
  privilege in the posix c run-time environment.
\newblock In {\em ASPLOS}, 2019.

\bibitem{imagenet}
J.~Deng, W.~Dong, R.~Socher, L.-J. Li, K.~Li, and L.~Fei-Fei.
\newblock {ImageNet: A Large-Scale Hierarchical Image Database}.
\newblock In {\em CVPR09}, 2009.

\bibitem{m2r}
Tien Tuan~Anh Dinh, Prateek Saxena, Ee-Chien Chang, Beng~Chin Ooi, and Chunwang
  Zhang.
\newblock {M2R: Enabling Stronger Privacy in MapReduce Computation}.
\newblock In {\em USENIX Security}, 2015.

\bibitem{iso-x}
D.~{Evtyushkin}, J.~{Elwell}, M.~{Ozsoy}, D.~{Ponomarev}, N.~A. {Ghazaleh}, and
  R.~{Riley}.
\newblock Iso-x: A flexible architecture for hardware-managed isolated
  execution.
\newblock In {\em MICRO}, 2014.

\bibitem{komodo}
Andrew Ferraiuolo, Andrew Baumann, Chris Hawblitzel, and Bryan Parno.
\newblock Komodo: Using verification to disentangle secure-enclave hardware
  from software.
\newblock In {\em SOSP}, 2017.

\bibitem{vale-popl}
Aymeric Fromherz, Nick Giannarakis, Chris Hawblitzel, Bryan Parno, Aseem
  Rastogi, and Nikhil Swamy.
\newblock A verified, efficient embedding of a verifiable assembly language.
\newblock In {\em POPL}, 2019.

\bibitem{terra}
Tal Garfinkel, Ben Pfaff, Jim Chow, Mendel Rosenblum, and Dan Boneh.
\newblock {Terra: A Virtual Machine-based Platform for Trusted Computing}.
\newblock In {\em {SOSP}}, 2003.

\bibitem{ostia}
Tal Garfinkel, Ben Pfaff, and Mendel Rosenblum.
\newblock {Ostia: A Delegating Architecture for Secure System Call
  Interposition}.
\newblock In {\em NDSS}, 2003.

\bibitem{flexible-os}
Tal Garfinkel, Mendel Rosenblum, and Dan Boneh.
\newblock Flexible os support and applications for trusted computing.
\newblock In {\em HOTOS}, 2003.

\bibitem{time-sc-survey}
Qian Ge, Yuval Yarom, David Cock, and Gernot Heiser.
\newblock A survey of microarchitectural timing attacks and countermeasures on
  contemporary hardware.
\newblock {\em Journal of Cryptographic Engineering}, 2018.

\bibitem{endbox}
David Goltzsche, Signe R{\"{u}}sch, Manuel Nieke, S{\'{e}}bastien Vaucher, Nico
  Weichbrodt, Valerio Schiavoni, Pierre{-}Louis Aublin, Paolo Costa, Christof
  Fetzer, Pascal Felber, Peter~R. Pietzuch, and R{\"{u}}diger Kapitza.
\newblock Endbox: Scalable middlebox functions using client-side trusted
  execution.
\newblock In {\em {DSN}}, 2018.

\bibitem{vif}
Deli Gong, Muoi Tran, Shweta Shinde, Hao Jin, Vyas Sekar, Prateek Saxena, and
  Min~Suk Kang.
\newblock {Practical Verifiable In-network Filtering for DDoS defense}.
\newblock In {\em ICDCS}, 2019.

\bibitem{boom-speculative}
Abraham Gonzalez, Ben Korpan, Jerry Zhao, Ed~Younis, and Krste Asanović.
\newblock {Replicating and Mitigating Spectre Attacks on a Open Source RISC-V
  Microarchitecture}.
\newblock In {\em Third Workshop on Computer Architecture Research with RISC-V
  (CARRV)}, 2019.

\bibitem{GrahamDenning1972}
G.~Scott Graham and Peter~J. Denning.
\newblock Protection: Principles and practice.
\newblock In {\em Spring Joint Computer Conference}, AFIPS, 1972.

\bibitem{certikos}
Ronghui Gu, Zhong Shao, Hao Chen, Xiongnan Wu, Jieung Kim, Vilhelm Sj\"{o}berg,
  and David Costanzo.
\newblock Certikos: An extensible architecture for building certified
  concurrent os kernels.
\newblock OSDI'16.

\bibitem{general-purpose-mee}
Shay Gueron.
\newblock A memory encryption engine suitable for general purpose processors.
\newblock Cryptology ePrint Archive, Report 2016/204, 2016.

\bibitem{ironclad}
Chris Hawblitzel, Jon Howell, Jacob~R. Lorch, Arjun Narayan, Bryan Parno,
  Danfeng Zhang, and Brian Zill.
\newblock Ironclad apps: End-to-end security via automated full-system
  verification.
\newblock In {\em {USENIX} {OSDI}}, 2014.

\bibitem{inktag}
Owen~S. Hofmann, Sangman Kim, Alan~M. Dunn, Michael~Z. Lee, and Emmett Witchel.
\newblock {InkTag: Secure Applications on an Untrusted Operating System}.
\newblock In {\em ASPLOS}, 2013.

\bibitem{PKI-CRL}
R.~Housley, W.~Polk, W.~Ford, and D.~Solo.
\newblock Internet x.509 public key infrastructure certificate and certificate
  revocation list (crl) profile, 2002.

\bibitem{seven-properties-highly-secure-devices}
Galen Hunt, George Letey, and Ed~Nightingale.
\newblock The seven properties of highly secure devices.
\newblock Technical report, March 2017.

\bibitem{intel-attestation-service}
Simon Johnson, Vinnie Scarlata, Carlos Rozas, Ernie Brickell, and Frank Mckeen.
\newblock Intel software guard extensions: Epid provisioning and attestation
  services, 2016.

\bibitem{amd-sev-es}
David Kaplan.
\newblock {Protecting VM Register State with SEV-ES}.
\newblock
  \url{http://support.amd.com/TechDocs/Protecting%20VM%20Register%20State%20with%20SEV-ES.pdf},
  2017.

\bibitem{amd-sev}
David Kaplan, Jeremy Powell, and Tom Woller.
\newblock {AMD Memory Encryption}.
\newblock
  \url{http://amd-dev.wpengine.netdna-cdn.com/wordpress/media/2013/12/AMD_Memory_Encryption_Whitepaper_v7-Public.pdf},
  2016.

\bibitem{firesim}
Sagar Karandikar, Howard Mao, Donggyu Kim, David Biancolin, Alon Amid, Dayeol
  Lee, Nathan Pemberton, Emmanuel Amaro, Colin Schmidt, Aditya Chopra, Qijing
  Huang, Kyle Kovacs, Borivoje Nikolic, Randy Katz, Jonathan Bachrach, and
  Krste Asanovi\'{c}.
\newblock Firesim: Fpga-accelerated cycle-exact scale-out system simulation in
  the public cloud.
\newblock In {\em ISCA}, 2018.

\bibitem{nohype}
Eric Keller, Jakub Szefer, Jennifer Rexford, and Ruby~B. Lee.
\newblock {NoHype: Virtualized Cloud Infrastructure without the
  Virtualization}.
\newblock In {\em {Proceedings of International Symposium on Computer
  Architecture}}, 2010.

\bibitem{microsemi-secure-boot}
Pierre~Selwan Ken~Irving.
\newblock Revolutionizing the computing landscape and beyond.
\newblock
  \url{https://content.riscv.org/wp-content/uploads/2018/12/RISC-V-MultiCore-Secure-Boot-Ken-Irvining-and-Pierre-Selwan.pdf},
  December 2018.

\bibitem{dawg}
Vladimir Kiriansky, Ilia Lebedev, Saman Amarasinghe, Srinivas Devadas, and Joel
  Emer.
\newblock Dawg: A defense against cache timing attacks in speculative execution
  processors.
\newblock In {\em MICRO}, 2018.

\bibitem{sel4}
Gerwin Klein, Kevin Elphinstone, Gernot Heiser, June Andronick, David Cock,
  Philip Derrin, Dhammika Elkaduwe, Kai Engelhardt, Rafal Kolanski, Michael
  Norrish, Thomas Sewell, Harvey Tuch, and Simon Winwood.
\newblock sel4: Formal verification of an os kernel.
\newblock In {\em SOSP}, 2009.

\bibitem{spectre}
Paul Kocher, Daniel Genkin, Daniel Gruss, Werner Haas, Mike Hamburg, Moritz
  Lipp, Stefan Mangard, Thomas Prescher, Michael Schwarz, and Yuval Yarom.
\newblock Spectre attacks: Exploiting speculative execution.
\newblock 2018.

\bibitem{trustlite}
Patrick Koeberl, Steffen Schulz, Ahmad-Reza Sadeghi, and Vijay Varadharajan.
\newblock Trustlite: A security architecture for tiny embedded devices.
\newblock In {\em EuroSys}, 2014.

\bibitem{sanctum-secure-boot}
Ilia Lebedev, Kyle Hogan, and Srinivas Devadas.
\newblock {Secure Boot and Remote Attestation in the Sanctum Processor}.
\newblock In {\em CSF}, 2018.

\bibitem{sanctorum}
Ilia~A. Lebedev, Kyle Hogan, Jules Drean, David Kohlbrenner, Dayeol Lee, Krste
  Asanovic, Dawn Song, and Srinivas Devadas.
\newblock Sanctorum: {A} lightweight security monitor for secure enclaves.
\newblock {\em {IACR} Cryptology ePrint Archive}, 2019.

\bibitem{sp}
Ruby~B. Lee, Peter C.~S. Kwan, John~P. McGregor, Jeffrey Dwoskin, and Zhenghong
  Wang.
\newblock {Architecture for Protecting Critical Secrets in Microprocessors}.
\newblock {\em SIGARCH Comput. Archit. News}, 2005.

\bibitem{compcert-paper}
Xavier Leroy.
\newblock Formal verification of a realistic compiler.

\bibitem{droidvault}
X.~{Li}, H.~{Hu}, G.~{Bai}, Y.~{Jia}, Z.~{Liang}, and P.~{Saxena}.
\newblock Droidvault: A trusted data vault for android devices.
\newblock In {\em ICECCS}, 2014.

\bibitem{teechain}
Joshua Lind, Ittay Eyal, Florian Kelbert, Oded Naor, Peter~R. Pietzuch, and
  Emin~G{\"{u}}n Sirer.
\newblock {Teechain: Scalable Blockchain Payments using Trusted Execution
  Environments}.
\newblock {\em CoRR}, abs/1707.05454, 2017.

\bibitem{trustvisor}
Jonathan~M. McCune, Yanlin Li, Ning Qu, Zongwei Zhou, Anupam Datta, Virgil
  Gligor, and Adrian Perrig.
\newblock {TrustVisor: Efficient TCB Reduction and Attestation}.
\newblock In {\em IEEE S\&P}, 2010.

\bibitem{flicker}
Jonathan~M. McCune, Bryan~J. Parno, Adrian Perrig, Michael~K. Reiter, and
  Hiroshi Isozaki.
\newblock {Flicker: An Execution Infrastructure for TCB Minimization}.
\newblock {\em SIGOPS Oper. Syst. Rev.}, 2008.

\bibitem{sgx-dyn-mem}
Frank McKeen, Ilya Alexandrovich, Ittai Anati, Dror Caspi, Simon Johnson,
  Rebekah Leslie-Hurd, and Carlos Rozas.
\newblock Intel software guard extensions support for dynamic memory management
  inside an enclave.
\newblock In {\em HASP}, 2016.

\bibitem{sgx}
Frank McKeen, Ilya Alexandrovich, Alex Berenzon, Carlos~V. Rozas, Hisham Shafi,
  Vedvyas Shanbhogue, and Uday~R. Savagaonkar.
\newblock {Innovative Instructions and Software Model for Isolated Execution}.
\newblock In {\em HASP}, 2013.

\bibitem{merkletree}
Ralph~C Merkle.
\newblock A digital signature based on a conventional encryption function.
\newblock In {\em Conference on the theory and application of cryptographic
  techniques}, pages 369--378. Springer, 1987.

\bibitem{proof-of-luck}
Mitar Milutinovic, Warren He, Howard Wu, and Maxinder Kanwal.
\newblock Proof of luck: an efficient blockchain consensus protocol.
\newblock {\em CoRR}, abs/1703.05435, 2017.

\bibitem{fallout}
Marina Minkin, Daniel Moghimi, Moritz Lipp, Michael Schwarz, Jo~Van~Bulck,
  Daniel Genkin, Daniel Gruss, Berk Sunar, Frank Piessens, and Yuval Yarom.
\newblock {Fallout}: Reading kernel writes from user space.
\newblock 2019.

\bibitem{mowery2013welcome}
Keaton Mowery, Michael Wei, David Kohlbrenner, Hovav Shacham, and Steven
  Swanson.
\newblock Welcome to the entropics: Boot-time entropy in embedded devices.
\newblock In {\em IEEE S\&P}, 2013.

\bibitem{asylo}
Jason~Garms Nelly~Porter.
\newblock Advancing confidential computing with asylo and the confidential
  computing challenge.
\newblock
  \url{https://cloud.google.com/blog/products/identity-security/advancing-confidential-computing-with-asylo-and-the-confidential-computing-challenge},
  feb 2019.

\bibitem{serval}
Luke Nelson, James Bornholt, Ronghui Gu, Andrew Baumann, Emina Torlak, and
  Xi~Wang.
\newblock {Serval: Scaling Symbolic Evaluation for Automated Verification of
  Systems Code}.
\newblock In {\em SOSP}, 2019.

\bibitem{hyperkernel}
Luke Nelson, Helgi Sigurbjarnarson, Kaiyuan Zhang, Dylan Johnson, James
  Bornholt, Emina Torlak, and Xi~Wang.
\newblock Hyperkernel: Push-button verification of an os kernel.
\newblock SOSP '17.

\bibitem{intelCAT}
Khang~T Nguyen.
\newblock {Introduction to Cache Allocation Technology in the Intel® Xeon®
  Processor E5 v4 Family}.
\newblock
  \url{https://software.intel.com/en-us/articles/introduction-to-cache-allocation-technology},
  Feb 2016.

\bibitem{sancus}
Job Noorman, Pieter Agten, Wilfried Daniels, Raoul Strackx, Anthony
  Van~Herrewege, Christophe Huygens, Bart Preneel, Ingrid Verbauwhede, and
  Frank Piessens.
\newblock Sancus: Low-cost trustworthy extensible networked devices with a
  zero-software trusted computing base.
\newblock In {\em USENIX Security}, 2013.

\bibitem{sgx-ml}
Olga Ohrimenko, Felix Schuster, Cedric Fournet, Aastha Mehta, Sebastian
  Nowozin, Kapil Vaswani, and Manuel Costa.
\newblock Oblivious multi-party machine learning on trusted processors.
\newblock In {\em USENIX Security}, 2016.

\bibitem{varys}
Oleksii Oleksenko, Bohdan Trach, Robert Krahn, Mark Silberstein, and Christof
  Fetzer.
\newblock Varys: Protecting {SGX} enclaves from practical side-channel attacks.
\newblock In {\em {USENIX} {ATC}}, 2018.

\bibitem{eleos}
Meni Orenbach, Pavel Lifshits, Marina Minkin, and Mark Silberstein.
\newblock Eleos: Exitless os services for sgx enclaves.
\newblock EuroSys, 2017.

\bibitem{cosmix}
Meni Orenbach, Yan Michalevsky, Christof Fetzer, and Mark Silberstein.
\newblock Cosmix: A compiler-based system for secure memory instrumentation and
  execution in enclaves.
\newblock In {\em {USENIX} {ATC}}, 2019.

\bibitem{oasis}
Emmanuel Owusu, Jorge Guajardo, Jonathan McCune, Jim Newsome, Adrian Perrig,
  and Amit Vasudevan.
\newblock Oasis: On achieving a sanctuary for integrity and secrecy on
  untrusted platforms.
\newblock In {\em CCS}, 2013.

\bibitem{sifive-tee}
Nate~Graff Palmer~Dabbelt.
\newblock {SiFive's Trusted Execution Reference Platform}.
\newblock
  \url{https://content.riscv.org/wp-content/uploads/2018/12/SiFives-Trusted-Execution-Reference-Platform-Palmer-Dabbelt-1-1.pdf},
  December 2018.

\bibitem{bootstrap-trust}
Bryan Parno, Jonathan~M. McCune, and Adrian Perrig.
\newblock Bootstrapping trust in commodity computers.
\newblock In {\em IEEE S\&P}, 2010.

\bibitem{drawbridge}
Donald~E. Porter, Silas Boyd-Wickizer, Jon Howell, Reuben Olinsky, and Galen~C.
  Hunt.
\newblock {Rethinking the Library OS from the Top Down}.
\newblock In {\em ASPLOS}, 2011.

\bibitem{ports-garfinkel}
Dan R.~K. Ports and Tal Garfinkel.
\newblock {Towards Application Security on Untrusted Operating Systems}.
\newblock In {\em HOTSEC}, 2008.

\bibitem{enclavedb}
Christian Priebe, Kapil Vaswani, and Manuel Costa.
\newblock Enclavedb - a secure database using sgx.
\newblock In {\em IEE S\&P}, 2018.

\bibitem{evercrypt}
Jonathan Protzenko, Bryan Parno, Aymeric Fromherz, Chris Hawblitzel, Marina
  Polubelova, Karthikeyan Bhargavan, Benjamin Beurdouche, Joonwon Choi, Antoine
  Delignat-Lavaud, Cedric Fournet, Tahina Ramananandro, Aseem Rastogi, Nikhil
  Swamy, Christoph Wintersteiger, and Santiago Zanella-Beguelin.
\newblock Evercrypt: A fast, verified, cross-platform cryptographic provider.
\newblock Technical Report 2019/757, ePrint, July 2019.

\bibitem{secureblue}
{Rick Boivie}.
\newblock {SecureBlue++: CPU Support for Secure Execution}.
\newblock 2012.

\bibitem{bonsaitree}
B.~{Rogers}, S.~{Chhabra}, M.~{Prvulovic}, and Y.~{Solihin}.
\newblock Using address independent seed encryption and bonsai merkle trees to
  make secure processors os- and performance-friendly.
\newblock In {\em MICRO 2007}, 2007.

\bibitem{aise}
Brian Rogers, Siddhartha Chhabra, Milos Prvulovic, and Yan Solihin.
\newblock {Using Address Independent Seed Encryption and Bonsai Merkle Trees to
  Make Secure Processors OS- and Performance-Friendly}.
\newblock In {\em MICRO}, 2007.

\bibitem{timberv}
{Samuel Weiser and Mario Werner and Ferdinand Brasser and Maja Malenko and
  Stefan Mangard and Ahmad-Reza Sadeghi}.
\newblock {TIMBER-V: Tag-Isolated Memory Bringing Fine-grained Enclaves to
  RISC-V}.
\newblock In {\em NDSS}, 2019.

\bibitem{vc3}
Felix Schuster, Manuel Costa, Cedric Fournet, Christos Gkantsidis, Marcus
  Peinado, Gloria Mainar-Ruiz, and Mark Russinovich.
\newblock {VC3: Trustworthy Data Analytics in the Cloud}.
\newblock In {\em IEEE S\&P}, 2015.

\bibitem{zombieload}
Michael Schwarz, Moritz Lipp, Daniel Moghimi, Jo~Van~Bulck, Julian Stecklina,
  Thomas Prescher, and Daniel Gruss.
\newblock {ZombieLoad}: Cross-privilege-boundary data sampling.
\newblock {\em arXiv:1905.05726}, 2019.

\bibitem{pigeonhole}
Shweta Shinde, Zheng~Leong Chua, Viswesh Narayanan, and Prateek Saxena.
\newblock Preventing page faults from telling your secrets.
\newblock ASIACCS'16.

\bibitem{panoply}
Shweta Shinde, Dat~Le Tien, Shruti Tople, and Prateek Saxena.
\newblock {Panoply: Low-TCB Linux Applications With SGX Enclaves}.
\newblock In {\em {NDSS}}, 2017.

\bibitem{besfs}
Shweta {Shinde}, Shengi {Wang}, Pinghai {Yuan}, Aquinas {Hobor}, Abhik
  {Roychoudhury}, and Prateek {Saxena}.
\newblock {BesFS: Mechanized Proof of an Iago-Safe Filesystem for Enclaves}.
\newblock {\em ArXiv e-prints}, July 2018.

\bibitem{nickel}
Helgi Sigurbjarnarson, Luke Nelson, Bruno Castro-Karney, James Bornholt, Emina
  Torlak, and Xi~Wang.
\newblock Nickel: A framework for design and verification of information flow
  control systems.
\newblock In {\em OSDI}, 2018.

\bibitem{msr-pldi16}
Rohit Sinha, Manuel Costa, Akash Lal, Nuno Lopes, Sanjit Seshia, Sriram
  Rajamani, and Kapil Vaswani.
\newblock A design and verification methodology for secure isolated regions.
\newblock In {\em PLDI '16}.

\bibitem{moat}
Rohit Sinha, Sriram Rajamani, Sanjit Seshia, and Kapil Vaswani.
\newblock Moat: Verifying confidentiality of enclave programs.
\newblock CCS '15.

\bibitem{tap}
Pramod Subramanyan, Rohit Sinha, Ilia Lebedev, Srinivas Devadas, and Sanjit~A.
  Seshia.
\newblock A formal foundation for secure remote execution of enclaves.
\newblock CCS '17.

\bibitem{aegis}
G.~Edward Suh, Charles~W. O'Donnell, Ishan Sachdev, and Srinivas Devadas.
\newblock {Design and Implementation of the AEGIS Single-Chip Secure Processor
  Using Physical Random Functions}.
\newblock {\em SIGARCH Comput. Archit. News}, 2005.

\bibitem{hyperwall}
Jakub Szefer and Ruby~B. Lee.
\newblock {Architectural Support for Hypervisor-secure Virtualization}.
\newblock In {\em ASPLOS}, 2012.

\bibitem{xom}
David Lie~Chandramohan Thekkath, Mark Mitchell, Patrick Lincoln, Dan Boneh,
  John Mitchell, and Mark Horowitz.
\newblock {Architectural Support for Copy and Tamper Resistant Software}.
\newblock In {\em ASPLOS}, 2000.

\bibitem{privado}
Shruti {Tople}, Karan {Grover}, Shweta {Shinde}, Ranjita {Bhagwan}, and
  Ramachandran {Ramjee}.
\newblock {Privado: Practical and Secure DNN Inference}.
\newblock {\em ArXiv}, 2018.

\bibitem{ridl}
Stephan van Schaik, Alyssa Milburn, Sebastian Österlund, Pietro Frigo, Giorgi
  Maisuradze, Kaveh Razavi, Herbert Bos, and Cristiano Giuffrida.
\newblock {RIDL}: Rogue in-flight data load.
\newblock In {\em S\&{P}}, 2019.

\bibitem{hyperapps}
Amit Vasudevan, Sagar Chaki, Limin Jia, Jonathan McCune, James Newsome, and
  Anupam Datta.
\newblock {Design, Implementation and Verification of an eXtensible and Modular
  Hypervisor Framework}.
\newblock In {\em IEEE S\&P}, 2013.

\bibitem{cheri}
R.~N.~M. {Watson}, J.~{Woodruff}, P.~G. {Neumann}, S.~W. {Moore},
  J.~{Anderson}, D.~{Chisnall}, N.~{Dave}, B.~{Davis}, K.~{Gudka}, B.~{Laurie},
  S.~J. {Murdoch}, R.~{Norton}, M.~{Roe}, S.~{Son}, and M.~{Vadera}.
\newblock Cheri: A hybrid capability-system architecture for scalable software
  compartmentalization.
\newblock In {\em IEEE Symposium on Security and Privacy}, 2015.

\bibitem{cca-sgx}
Yuanzhong Xu, Weidong Cui, and Marcus Peinado.
\newblock {Controlled-Channel Attacks: Deterministic Side Channels for
  Untrusted Operating Systems}.
\newblock In {\em IEEE S\&P}, 2015.

\bibitem{invisispec}
M.~{Yan}, J.~{Choi}, D.~{Skarlatos}, A.~{Morrison}, C.~{Fletcher}, and
  J.~{Torrellas}.
\newblock Invisispec: Making speculative execution invisible in the cache
  hierarchy.
\newblock In {\em MICRO}, 2018.

\bibitem{verve}
Jean Yang and Chris Hawblitzel.
\newblock Safe to the last instruction: Automated verification of a type-safe
  operating system.
\newblock In {\em PLDI}, 2010.

\bibitem{sp3}
Jisoo Yang and Kang~G. Shin.
\newblock {Using Hypervisor to Provide Data Secrecy for User Applications on a
  Per-page Basis}.
\newblock In {\em VEE}, 2008.

\bibitem{towncrier}
Fan Zhang, Ethan Cecchetti, Kyle Croman, Ari Juels, and Elaine Shi.
\newblock Town crier: An authenticated data feed for smart contracts.
\newblock In {\em CCS}, 2016.

\bibitem{cloudvisor}
Fengzhe Zhang, Jin Chen, Haibo Chen, and Binyu Zang.
\newblock {CloudVisor: Retrofitting Protection of Virtual Machines in
  Multi-Tenant Cloud with Nested Virtualization}.
\newblock In {\em SOSP}, 2011.

\end{thebibliography}

\end{document}